\documentclass[showpacs,prb,superscriptaddress,twocolumn]{revtex4-1}
\usepackage[table]{xcolor}
\usepackage{textcomp}
\usepackage{amsmath}
\usepackage{amssymb}
\usepackage{}
\usepackage{graphicx} 
\usepackage{}
\usepackage{pgf}
\usepackage{epstopdf}
\usepackage[pdftex,bookmarks,colorlinks]{hyperref}
\usepackage{hyperref}

\newcommand{\ket}[1]{\left|#1\right\rangle}
\newcommand{\bra}[1]{\left\langle #1\right|}

\newcommand{\mean}[1]{\left\langle #1\right\rangle}

\newcommand{\nep}{\textrm{e}}

\newcommand{\Ham}{\widehat{H}}

\newcommand{\calH}{\mathcal{H}}

\newcommand{\opS}{\widehat{S}}

\newcommand{\red}[1]{\textcolor{black}{#1}}

\begin{document}
\title{Floquet time crystal in the Lipkin-Meshkov-Glick model}

\author{Angelo Russomanno}
\affiliation{NEST, Scuola Normale Superiore \& Istituto Nanoscienze-CNR, I-56126 Pisa, Italy}
\affiliation{Abdus Salam ICTP, Strada Costiera 11, I-34151 Trieste, Italy}

\author{Fernando Iemini}
\affiliation{Abdus Salam ICTP, Strada Costiera 11, I-34151 Trieste, Italy}

\author{Marcello Dalmonte}
\affiliation{Abdus Salam ICTP, Strada Costiera 11, I-34151 Trieste, Italy}
\author{Rosario Fazio}
\affiliation{Abdus Salam ICTP, Strada Costiera 11, I-34151 Trieste, Italy}
\affiliation{NEST, Scuola Normale Superiore \& Istituto Nanoscienze-CNR, I-56126 Pisa, Italy}

\begin{abstract}
In this work we discuss the existence of time-translation symmetry breaking in a kicked \red{infinite-range-interacting} clean spin system described by the 
Lipkin-Meshkov-Glick model. This Floquet time crystal is robust under perturbations of the kicking protocol, its existence being intimately linked to 
the underlying $\mathbb{Z}_2$  symmetry breaking of the time-independent model. We show that the model being \red{infinite-range} and having an 
extensive amount of symmetry breaking eigenstates is essential for having the time-crystal behaviour. In particular we discuss the 
properties of the Floquet spectrum, and show the existence of doublets of Floquet states which are respectively even and odd superposition 
of symmetry broken states and have quasi-energies differing of half the driving frequencies, a key essence of Floquet time crystals. 
{Remarkably, the stability of the time-crystal phase can be directly analysed in the limit of infinite size, discussing the properties of the corresponding classical phase space. }Through a detailed analysis of the robustness of the time crystal to various perturbations we are able to map
the corresponding phase diagram. 
We finally discuss  the possibility of an experimental implementation by means of trapped ions.
\end{abstract}

\maketitle
\section{Introduction}
Landau's idea of classifying phases of matter in terms of symmetry breaking  is a cornerstone of modern physics~\cite{Goldenfeld:book}. 
Breaking space translation symmetry gives rise to crystals, while superfluids and ferromagnets are manifestations of the spontaneous breaking 
of gauge or rotational invariance respectively.  Amongst all possible complex situations that were considered and experimentally verified so far, breaking 
the time-translation symmetry has received attention only {very recently.~\cite{wilczek_prl12,cinesi_prl12,wilczek_prl13,bruno_prl13a,bruno_prl13b,volovik_JETP13,pruno_prl13c,noziere_EPL13,Sacha_PRA15,Chandran_PRB15,Vedika2016,Nayak_PRL16,
Vedika_PRL16,Keyser_PRB16}} This was the  focus of the pioneering work of Wilczek~\cite{wilczek_prl12,wilczek_prl13} in which he argues that an autonomous system can break time translation symmetry, thus realising what he named as {\it time
crystals}. This possibility has been  ruled out~\cite{haruki:2014} for systems, with not too long range interactions, in their ground state or in thermal 
equilibrium~\cite{Note_S}.

The no-go theorems proved in Refs.~\onlinecite{haruki:2014,pruno_prl13c} clearly indicate that the right context where to search for spontaneous time-translation 
symmetry breaking is in systems out of equilibrium. The most fruitful setting so far has been provided by periodically driven systems. Since the 
pioneering {works~\cite{Nayak_PRL16,Vedika_PRL16,Keyser_PRB16} on {\it Floquet time crystals}~\cite{Nayak_PRL16} (a.k.a. $\pi$-spin glasses~\cite{Vedika_PRL16,Keyser_PRB16})}, the literature on the subject has vigorously 
flourished~\cite{choi_16:preprint,Vedika2016,Vedika_PRB16,zhang_16:preprint,yao:preprint}. All these proposals consider a many body system 
unitarily evolving under an external periodic driving with period $\tau$. The time translation symmetry breaking appears as the response of an observable 
which oscillates with periodicity that is a multiple of the imposed drive (in most cases it is period doubling $2\tau$). These "anomalous" oscillations persist
indefinitely, on approaching the thermodynamic limit. Only in this limit  time-translation symmetry breaking occurs, as it happens for any standard symmetry breaking. 
It is worth noting that time translation symmetry breaking 
{seems to be always strictly connected with a global symmetry breaking ($\mathbb{Z}_2$ 
symmetry in Refs.{~\cite{Vedika2016,Nayak_PRL16,Vedika_PRL16,Keyser_PRB16}}) leading to the  the concept of ``spatio-temporal order''~\cite{Vedika_PRL16,Vedika2016,Keyser_PRB16}}.

A  key element in characterising Floquet time crystals, at least in the initial works, was the presence of many-body localization (MBL).~\cite{Basko_Ann06,
Oganesyan_PRA75} In presence of disorder, \red{short-range} interacting spin or electron systems show no diffusion due to the existence of an extensive amount of 
local integrals  of motion (see e.g. the review Ref.~\onlinecite{Nandisko_ann15}). In the present context this property forbids the system to heat up, thus 
avoiding the destruction of the time crystal. Clean driven systems can be ergodic, and asymptotically reach infinite temperature~\cite{Rigol_PRX14,
Lazarides_PRE14,Ponte_AP15,Rosch_PRA15,Russomanno_EPL15} (a condition in which there is no time dependence and therefore no time-translation 
symmetry breaking), or be integrable and reach a time-periodic Generalised Gibbs ensemble 
which has the property of being periodic with period $\tau$~\cite{Russomanno_PRL12,Lazarides_PRL14,angelo_arXiv16}: again, no time-translation 
symmetry breaking. On the contrary, in MBL systems, the absence of diffusion forbids the excitations to propagate along the chain: in this way localised operators 
can be constructed whose Heisenberg dynamics leads to a period-doubling $2\tau$ (or more generally to a multiple of the imposed period). These operators 
give rise to the order parameter of the time-translation symmetry breaking~\cite{Vedika2016}. 

\red{The interplay between disorder and long range interactions was instead considered in Refs.~\onlinecite{wowei_17:preprint,choi_16:preprint,Smith_Natp12}. In these works, the authors 
take a disordered system with power-law interactions:  only for a given choice of parameters the time-crystal behaviour sets in.  There is however a regime of 
parameters such that the oscillations breaking the time-translation symmetry decay with a rate exponentially small in the deviation from the critical parameters (quasi-time-translation symmetry breaking).}

In this framework it came as a surprise the recent proposal of a system without disorder showing genuine time-translation symmetry breaking~\cite{tc_huang}. 
In this reference, the authors consider a 
ladder closely related to the Hubbard model: in this case the time-crystal behaviour is connected to the localisation due to the Hubbard interaction.

The aim of this paper is to present a second 
example of a clean system showing time-crystal behaviour: \red{the Lipkin-Meshkov-Glick model}. 
The main virtues of the model we consider are {\em i)} a 
truly time-crystal regime can be realised in a finite region of system parameters, and {\em ii)} 
{the Hamiltonian dynamics we discuss is immediately available with trapped ion experiments~\cite{blatt2012quantum,Jurcevic:2016aa,Neyenhuis:2016aa}}.
In the following we are considering a spin network with infinite range interactions described by the Lipkin-Meshkov-Glick model. The arguments of 
Ref.~\onlinecite{Vedika2016} on the absence of clean time crystals do not apply here. Being all the sites \red{interacting} with each other, the spreading of 
correlations does not imply that  the local observables cannot come back to themselves, after a given multiple period. 
{Due to the infinite range nature of our model, mean field turns to be exact in the thermodynamic limit. Time-translation symmetry breaking in periodically driven mean-field models has been also described in Refs.~\onlinecite{Sacha_PRA15} (an ultra-cold atomic cloud described by Gross-Pitaevskij equation) and~\onlinecite{Chandran_PRB15} (an $O(N)$ model with $N\to \infty$).} 

%
\red{ Some of the features related to the time-translation symmetry breaking in this model were found, for finite number of spins and a different form of driving, in Ref.~\onlinecite{PRA_Holthaus}. Here we further discuss the time-crystal properties in the light of the developments of Refs.~\onlinecite{Vedika2016,Nayak_PRL16,Vedika_PRL16,Keyser_PRB16}: we show how the time-translation symmetry breaking appears in the thermodynamic limit of infinite number of spins, we discuss its relation with the $\pi$-spectral pairing of the Floquet spectrum, its deep connection with the standard $\mathbb{Z}_2$ symmetry breaking and its robustness under modifications of the initial state and the driving parameters.}

The paper is organised as follows. In the next Section we introduce the Lipkin-Meshkov-Glick model (LMG) and its main properties together with the type of 
driven dynamics that will be considered in the rest of the paper. In Section~\ref{Time-crist:sec} we briefly review the definition of time crystal and list the 
observables that we will consider in order to characterise it. In Section~\ref{results:sec} we  show the existence of time-translation symmetry breaking oscillations persisting for an infinite time in the thermodynamic limit. We first consider an idealised case in Section~\ref{time_crist_pi:ssec}. We then discuss, in Section~\ref{rigido_come_rocco:sec},
that the behaviour we find is robust: the time crystal behaviour is observed in a whole range of parameters. In Section~\ref{Nayak:sec} we 
discuss the properties of the Floquet spectrum and find that they are in agreement with those found in {Refs.~\onlinecite{Nayak_PRL16,Vedika_PRL16,Keyser_PRB16}}. 
It is possible to understand the time-crystal behaviour  in the LMG  model in terms of the phase space properties of the classical model 
describing our system in the thermodynamic limit. As already mentioned, the situation we consider in the paper is amenable of an immediate experimental 
verification. In Section~\ref{experimentum:sec} we discuss the perspectives of experimental realisation. Section~\ref{concolina:sec} is devoted to
our conclusions and to a brief discussion of possible future directions. 

\section{The driven Lipkin-Meshkov-Glick model} 
\label{model:sec}
The system we consider in this work is an infinite-range spin model which is also defined as the LMG model~\cite{Lipkin_NucPhys65}. 
It can be experimentally realised in many ways: thanks to a mapping to an  interacting  two-component  Bose-Einstein  condensate (BEC)  with  linear  coupling~\cite{Obertallo:arXiv}, by a BEC in 
a double well~\cite{Albiez_PRL05} or using trapped ions~\cite{blatt2012quantum,Neyenhuis:2016aa,Jurcevic:2016aa}. 
We will consider in more detail the implementation with trapped ions 
 at the end of the paper. 

\begin{figure}[h!]
  \begin{center} 
    \begin{tabular}{c}
      \hspace{0cm}\resizebox{80mm}{!}{\includegraphics{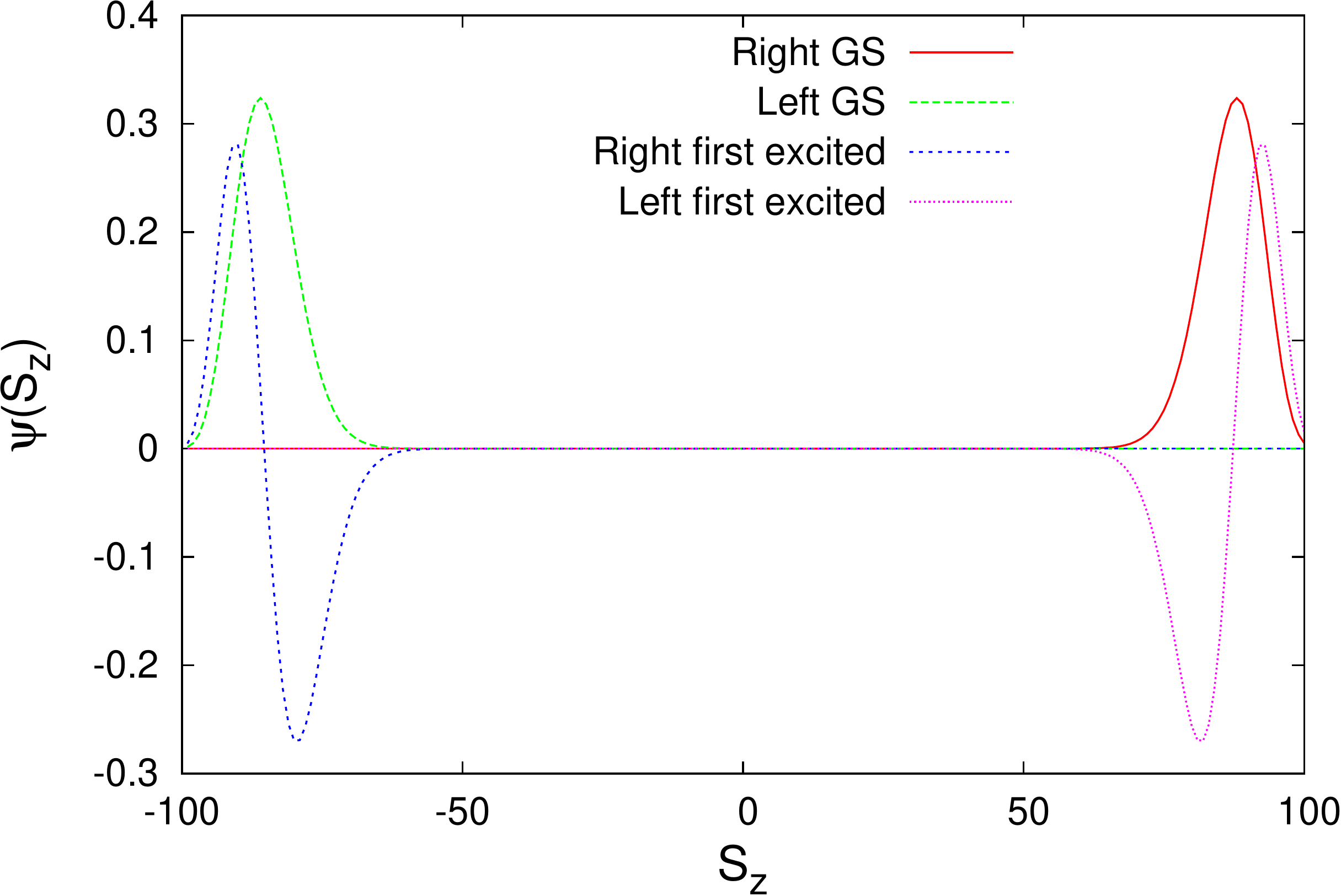}}\\
    \end{tabular}
  \end{center}
\caption{Some broken symmetry eigenstates of the Hamiltonian Eq.~\eqref{Lipkin-model:eqn} with $J=1$, $h<1$ and energies below the broken symmetry edge $E^*=-Nh$. We plot their (real) amplitudes in the $\widehat{S}^z$ representation. Numerical parameters: $N=200,\,h=0.5$.}
\label{broken:fig}
\end{figure}

The Hamiltonian of the LMG model is defined as
\begin{equation}  \label{Lipkin-model:eqn}
  \Ham(h)= -\frac{2J}{N}\sum_{i,j}^N \widehat{s}_{i}^z \widehat{s}_j^z - 2 h \, \sum_{i}^N \widehat{s}_i^{\,x} \;.
\end{equation}
It describes $N$ spin-1/2 ($\widehat{s}_{i}^{\alpha}$ is the $\alpha$-th component of the $i$-th spin) interacting through an infinite 
range coupling, in the presence of an external magnetic field $h$ along the $x-$direction.
This Hamiltonian conserves the total spin $\widehat{S}^2$  ($\widehat{S}^{\alpha}=\sum_{i}\widehat{s}_{i}^{\alpha}$), so we restrict to the spin sector with 
$S=N/2$ (we choose this particular value of $S$ because it is the one corresponding to the ground state~\cite{Mazza_PRB12}). When $h<J$ 
there is $\mathbb{Z}_2$ symmetry breaking that involves a finite fraction of all the spectrum. In the thermodynamic limit, $N\to \infty$, if we consider 
energy eigenvalues below the broken symmetry edge $E^*\equiv-h N$, the corresponding eigenstates appear in degenerate doublets. Each 
member of the pair is localised in the basis of the $\widehat{S}^z$-eigenstates $\ket{S_z}$, respectively at positive or negative values of the eigenvalue $S_z$ 
(see Fig.~\ref{broken:fig}). There is indeed an extensive fraction of the spectrum showing $\mathbb{Z}_2$ symmetry breaking. This kind of localisation occurs in the thermodynamic limit; for finite size, the true eigenstates are the even and 
odd superpositions of each doublet: the levels of each quasi-degenerate doublet are separated by a gap exponentially small in $N$. 

In all the text we consider $J=1$, so that there is the symmetry breaking phase for $h<1$.  

The properties of the LMG model have been extensively studied in the literature.  Recently  many works appeared concerning its non-equilibrium 
dynamics~\cite{Bapst_JSTAT12,Caneva_PRB08,Das_PRB06,Jorg_EPL10,Sciolla_JSTAT11,Mazza_PRB12}, especially in connection with a periodic 
driving~\cite{Russomanno_EPL15,Obertallo:arXiv,Haake_ZPB86,PRA_Holthaus}. Also in this work we want to consider a periodically driven dynamics that we specify 
in the following of this section. 
We first present and discuss the choice and the preparation of the initial states of our dynamics and then describe the driving protocol under
which we make them evolve.
\paragraph{Initial states} We initialize the system in one of the two symmetry breaking ground states of the Hamiltonian~\eqref{Lipkin-model:eqn},
with $h$ equal to some $h_{\rm i}<1$. For definiteness, let us consider the state with negative $z-$magnetization $\ket{\psi_{\rm GS}^-(h_{\rm i})}$. From a technical
point of view, to experimentally prepare this state also for $N$ finite, one must take the ground state of the Hamiltonian $\widehat{H}_{\rm i} = \widehat{H}(h_{\rm i})+ \delta h_z\sum_j\widehat{s}_j^z$ with $\delta h_z\ll 1$. The small field along $z$ breaks the $\mathbb{Z}_2$ symmetry and makes the ground state localized at negative values of $S_z$
(as in the thermodynamic limit) and no more even under this symmetry.
%
%
%
%
\paragraph{Driving protocol} After the initialisation the system will undergo a periodic driving dictated by the following time-evolution operator over one period
\begin{equation} \label{eckickazzo:eqn}
  \widehat{U}=\widehat{U}_{\rm kick}\exp\left[-i\widehat{H}(h)\tau\right]\;{\rm with}\;\widehat{U}_{\rm kick}\equiv\exp\left[-i{\phi}\sum_{i}^N \widehat{s}_i^{\,x}\right]\,,
\end{equation}
with $h<1$ (we will clarify in Section~\ref{results:sec} the reason behind this choice).
The system evolves with $\Ham(h)$; at times $t_n=n\tau$ the kicking operator $\widehat{U}_{\rm kick}$ acts as a rotation around the $x$ axis. 

We are going to analyse the long-time dynamics of the system 
as a function of $h$, $\phi$ and the choice of the initial state, for different values of $N$. We will show that there are regimes were the period-doubling appears in the 
thermodynamic limit thus confirming the existence of a time crystal in the LMG model. Before presenting the results, in the next Section we recap the salient 
features that enable us to characterise time-translation symmetry breaking.

\section{Observables in the time-crystal phase} \label{Time-crist:sec}
All quantum systems naturally show oscillations: this leads to phenomena which range from the Rabi to the Josephson oscillations; 
from the Bloch oscillations to the dynamical localisation. In order  to spot time-translation symmetry breaking, it is very important to 
define precise criteria which are able to distinguish this complex collective phenomenon from analogous single particle effects. A crucial step in this 
direction was done in {Refs.~\onlinecite{Nayak_PRL16,Vedika_PRL16,Keyser_PRB16}} where the relevant criteria and conditions to have a Floquet time crystal were introduced.
We do not attempt to recap here their formulation, the goal of this Section is to summarise the various indicators that will help us in identifying a time crystal 
regime in the LMG model (see also Ref.~\onlinecite{tc_huang}). 
 Following these previous works, \red{there must exist} 
an observable $\widehat{O}$ \red{(the order parameter)} and a class of initial states $\ket{\psi}$ such that, considering 
 stroboscopic times $t=n\tau$, \red{the expectation value in the thermodynamic limit ($N\to\infty$)}
\begin{equation} \label{thermol:eqn}
  f(t) = \lim_{N\to\infty}\bra{\psi(t)}\widehat{O}\ket{\psi(t)}
\end{equation}
satisfies all of the three conditions
\begin{itemize}
  \item[{\em I)}] Time-translation symmetry breaking: $f(t+\tau)\neq f(t)$ while $\widehat{H}(t+\tau)=\widehat{H}(t)$.
  \item[{\em II)}] Rigidity: $f(t)$ shows a fixed oscillation period $\tau_B$ (for instance $2\tau$) without fine-tuned Hamiltonian parameters.
  \item[{\em III)}] Persistence: the non-trivial oscillation with fixed period $\tau_B$ must persist for infinitely long time, when the \red{thermodynamic limit $N\to\infty$ in Eq.~\eqref{thermol:eqn} has been performed.} 
Thus, 
the Fourier transform $f_\omega$ must present a marked peak at the symmetry breaking frequency $\omega_B=2\pi/\tau_B$.
\end{itemize}
%

Furthermore it is important that the Floquet eigenstates -- the eigenstates of the stroboscopic dynamics -- have long range correlations~\cite{Nayak_PRL16}: in the case $\tau_B=2\tau$ we are focusing on, these states 
appear in pairs with quasi-energies (the corresponding eigenfrequencies) differing by an amount of $\pi/\tau$. In summary  we seek for an observable, the 
order parameter, such that it oscillates with frequency $2\tau$ for an infinite time in the thermodynamic limit (when the size of the system $N$ tends to infinity).
\red{The quantity we find to obey the three conditions listed above is the $z$-magnetization evaluated immediately before the $n$-th kick}
\begin{equation} \label{manna:eqn}
 m_n^z = \frac{1}{N}\bra{\psi(n\tau^-)}\widehat{S}^z\ket{\psi(n\tau^-)}\,,
\end{equation}
where $\ket{\psi(n\tau^-)}$ is the wave-function of the system just before the $n-$th period.
%
%

In the next sections we are going to show that this object meets all the three conditions.
%
%

\section{Results} 
\label{results:sec}

\begin{figure}
 \begin{center}
  \begin{tabular}{c}
    \resizebox{80mm}{!}{\includegraphics{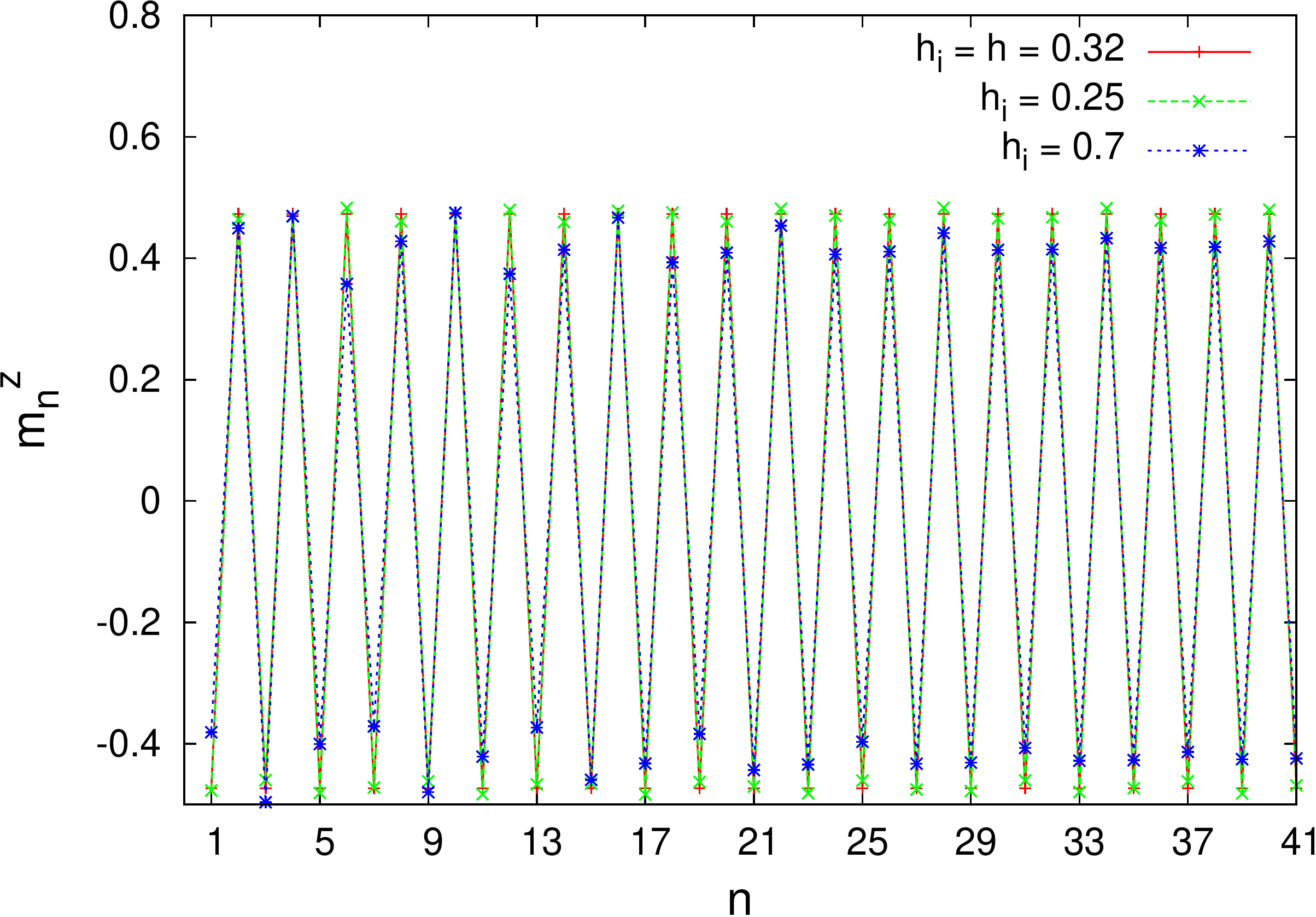}}\\
  \end{tabular}
  \caption{Stroboscopic magnetization ($t=n\tau$) for $N=100$ and kicking given by Eq.~\eqref{eckickazzo:eqn} with $h_{\rm i}=0.32$ and $\phi=\pi$. We see persistent oscillations with period $\tau_B=2\tau$ occurring for many different initial conditions $\ket{\psi_{\rm GS}^-(h_{\rm i})}$ (see the main text).}
\label{fig_clearest:fig}
  \end{center}
\end{figure}

In the rest of the paper we characterise in details the time-crystal behaviour of the driven LMG dynamics. In the same spirit of Ref.~\onlinecite{Nayak_PRL16} 
we first consider a simple situation.  

%
\subsection{Time crystal phase:  $\phi = \pi$} \label{time_crist_pi:ssec}

The picture is clearest when $\phi=\pi$ and $h_{\rm i}=h$ (the system is initialized in a symmetry breaking ground state of $\widehat{H}(h)$). In this case the kicking 
swaps the two degenerate symmetry-breaking ground states of $\widehat{H}(h)$ (that's why we need $h<1$, otherwise there is no ground state breaking the $\mathbb{Z}_2$ symmetry).  After the preparation in the negative magnetization symmetry-breaking ground state of $\widehat{H}(h)$, the system is commuted at each kick from the negative magnetization ground state to the positive one, and vice versa: 
it changes sign at each kick and gives rise to a period-doubling time crystal. We can see the persisting oscillations in Fig.~\ref{fig_clearest:fig},
where they appear for many different initial conditions. 

Fig.~\ref{fig_clearest:fig} shows the case of different initial $\ket{\psi_{\rm GS}^-(h_{\rm i})}$ with $h_{\rm i}<1$. 
Expressing each of these initial states in the basis of the eigenstates of $\widehat{H}(h)$, we have numerically checked -- for many values of $N$ -- that in this superposition there are only eigenstates of the symmetry breaking sector, whose number is extensive in $N$. All these eigenstates have 
energy below the broken symmetry edge and negative $z$-magnetization. 
Each of these states has a degenerate partner with positive magnetization: these pairs obeys the same qualitative picture that we have described for the ground state leading to time-translation symmetry breaking. This picture is still valid in the limit $N\to\infty$ thanks to the extensive number of symmetry-breaking eigenstates. This extensivity comes from the model being infinite range: this property of the interactions is therefore very important for the time-translation symmetry breaking. 
%
%
\red{We find indeed that our order parameter is the $z$-magnetization $\widehat{O}=\widehat{S}^z/N$: we are going to see how the time-translation symmetry breaking appears 
in the dynamics of this operator when we increase $N$ and approach the thermodynamic limit.}

\begin{figure}
  \begin{tabular}{c}
%
   \hspace{0cm}\resizebox{80mm}{!}{\includegraphics{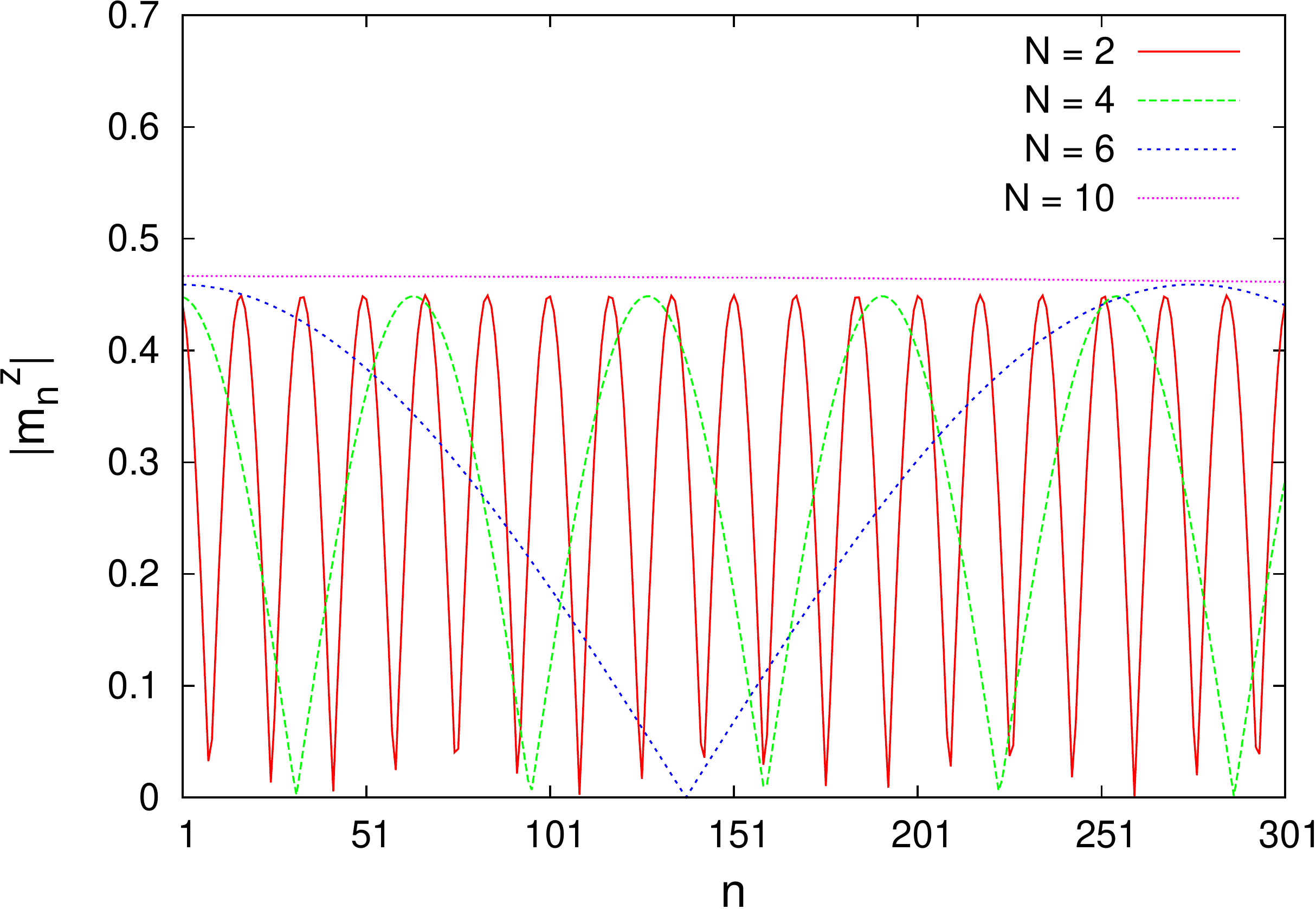}}
  \end{tabular}
  \caption{Beatings of $|m_n^z|$ for different values of $N$. Numerical parameters $h_{\rm i}=h=0.32,\,\phi=\pi,\,\tau=0.6$.}
  \label{fig_battimenti}
\end{figure}
Let us again start our analysis from the simpler case in which the initial state is $\ket{\psi_{\rm GS}^-(h)}$. When $N$ is finite the symmetric and antisymmetric ground states are non degenerate and 
there are some beatings superimposed to the period doubling oscillations (see Fig.~\ref{fig_battimenti} where we show $|m_n^z|$ 
for different values of $N$). 
\red{We note that, for finite $N$, the beatings are also present for a sinusoidal driving~\cite{PRA_Holthaus}.} We find that the period of the beatings is given by the inverse gap between the symmetric and antisymmetric ground states: the period is exponentially large in $N$ being the gap exponentially small.

\begin{figure}
  \begin{tabular}{c}
%
   \resizebox{80mm}{!}{\includegraphics{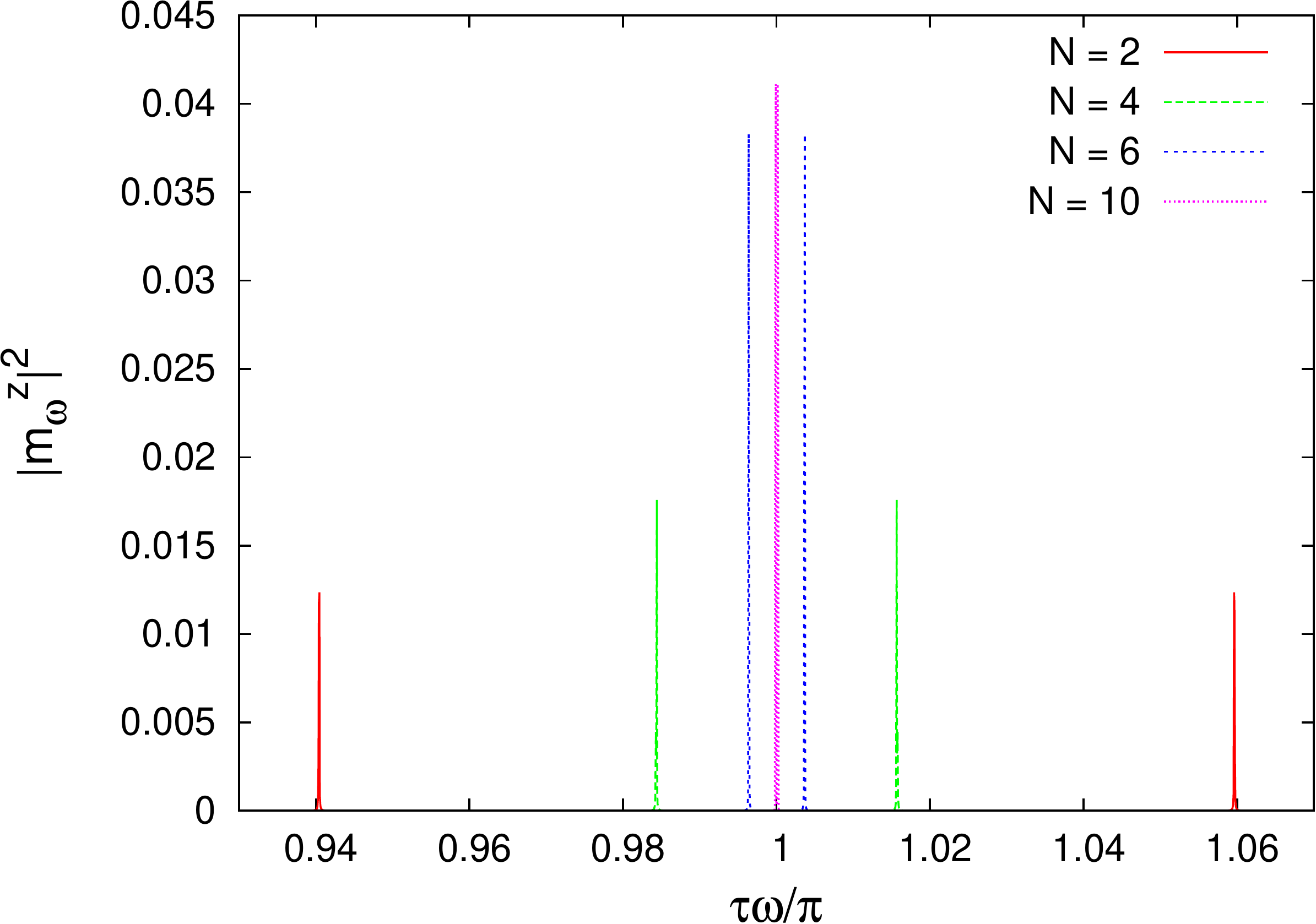}}\\
%
   \resizebox{80mm}{!}{\includegraphics{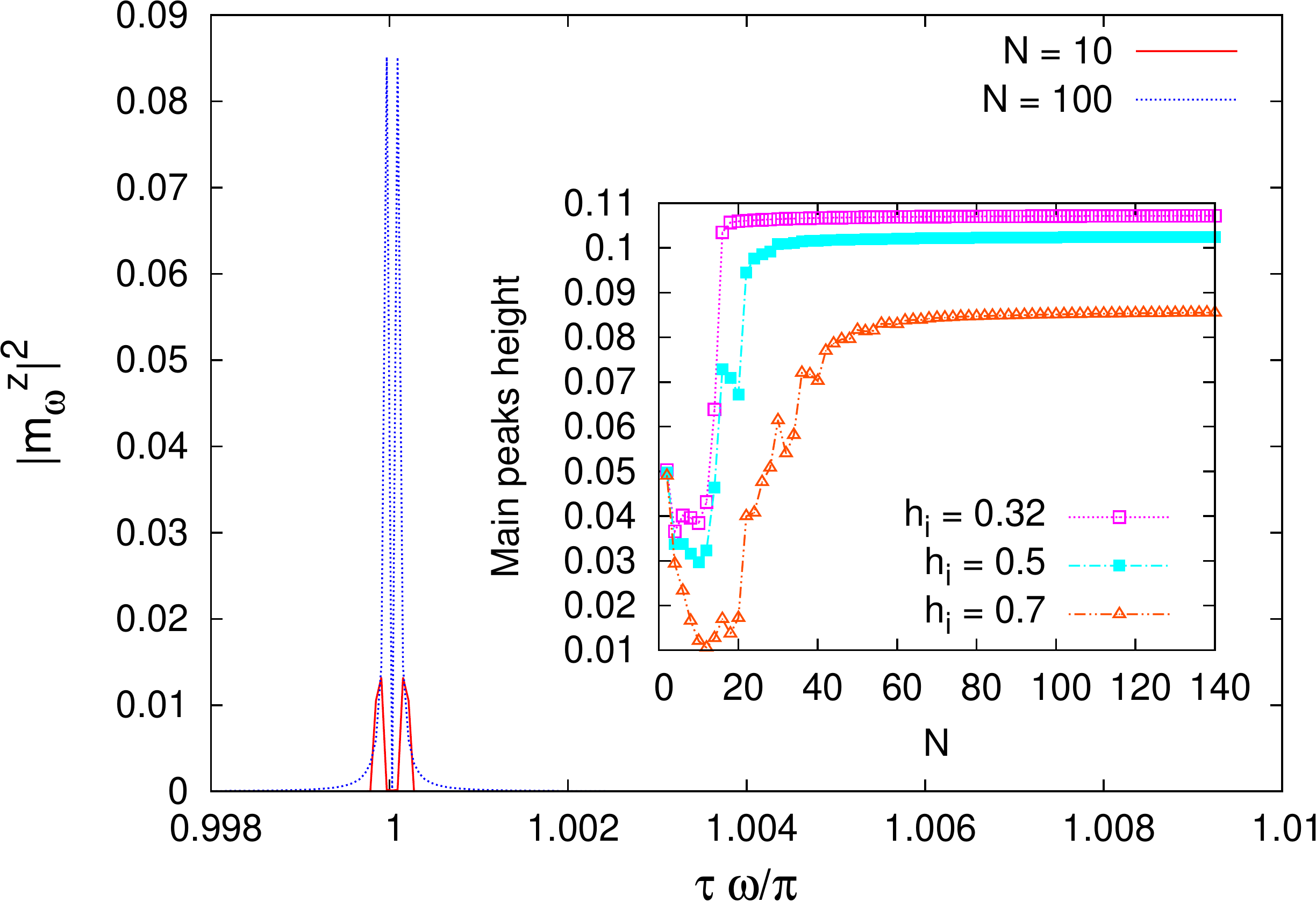}}\\
  \end{tabular}
  \caption{(In all the plots $\tau=0.6,\,\phi=\pi$.) Upper panel: power spectrum of the Fourier transform of $m_n^z$ for different values of $N$; notice the doublet of marked peaks around $\omega_B=\pi/\tau$ (here $h_{\rm i}=h=0.32$) which shrink into a single one as $N$ increases. Lower panel, main figure: power spectrum of the Fourier transform of $m_n^z$ for different $N$, $h_{\rm i}=0.7,\,h=0.32$. Notice the same peaks here. Lower panel, inset: plot of the height of the maximum peaks around $\omega=\pi/\tau$ vs $N$. It converges to a constant for $N\to\infty$: this highlights the existence of the time-crystal behaviour in the thermodynamic limit. (The Fourier transforms have been performed over 32768 driving periods in the main figures and over $K=65536$ periods in the insets.)}
  \label{fig1}
\end{figure}

We can see this phenomenon in $|m_\omega^z|^2$, the power spectrum of the $z$-magnetization discrete Fourier transform numerically performed over $K$ periods
\begin{equation} \label{transform:eqn}
  m_\omega^z = {\tau}\sum_{n=1}^{K}m_n^z\nep^{-i\omega n \tau}\quad{\rm with}\quad K \gg 1
\end{equation}
(upper panel of Fig.~\ref{fig1}). We see indeed in $|m_\omega^z|^2$ two peaks at 
$\omega_{\pm}(N)=\pi/\tau\pm\Delta(N)/2$ whose separation $\Delta(N)$ is the gap and exponentially decreases with $N$ as $\Delta(N)\sim\exp(-1.5 N)$. 
In the lower panel of Fig.~\ref{fig1} (main plot) we show $|m_\omega^z|^2$ vs $\omega$ in a case in which $h_{\rm i}\neq h$. 
We see that the same peak at $\omega_B=\pi/\tau$ emerges: it marks the existence of the time-translation 
symmetry breaking phase. The persistence of the oscillations for $N\to\infty$ can be seen in the inset of the lower panel of Fig.~\ref{fig1}. For different values of $h_{\rm i}$, we plot the height of the main peaks $|m_{\omega_{\pm}(N)}^z|^2$ 
vs $N$. 
$|m_{\omega_{\pm}(N)}^z|^2$ tends to a constant for $N\to\infty$: in the same limit $\omega_{\pm}(N)$ tend to $\omega_B$. Indeed, in the thermodynamic limit there are  persistent oscillations with period $\tau_B=2\tau$: the system breaks the time-translation symmetry. 
%

Conditions {\em I)} and {\em III)} for the existence of the time crystal are indeed fulfilled. We are going to  discuss condition {\em II)}
concerning the rigidity in the next Subsections.
%
%
\subsection{Robustness of the time crystal phase: $\phi \ne \pi$}
 \label{rigido_come_rocco:sec}
In this Subsection we want to show that the time crystal persists, with oscillations of $m_n^z$ rigidly fixed at period $\tau_B=2\tau$, for the phase $\phi$ of the kicking (see Eq.~\eqref{eckickazzo:eqn}) in a finite interval around $\pi$.

\begin{figure}
  \begin{tabular}{cc}
   \hspace{0cm}\resizebox{80mm}{!}{\includegraphics{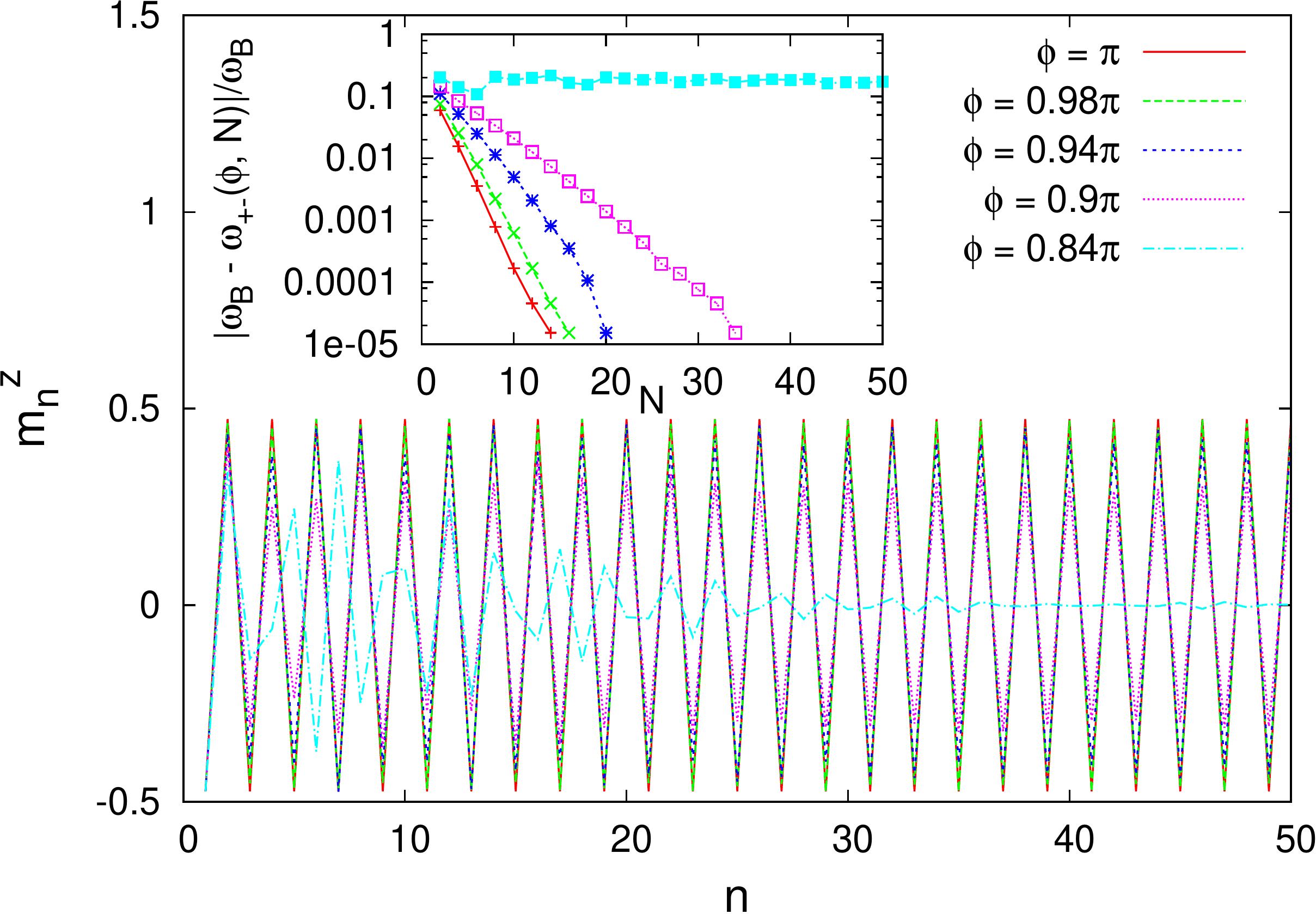}}\\
   \hspace{0cm}\resizebox{80mm}{!}{\includegraphics{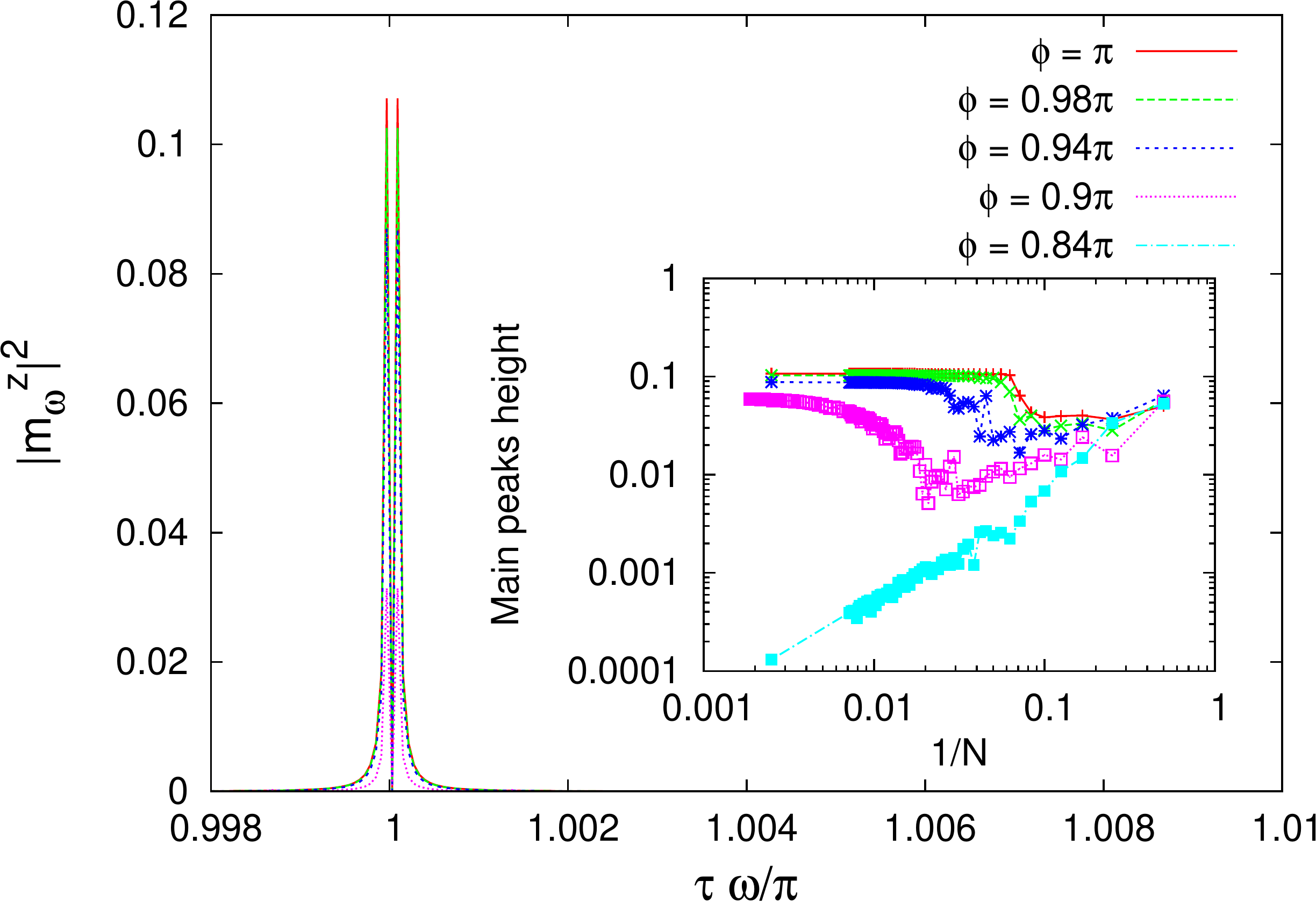}}\\
%
  \end{tabular}
  \caption{{\em Main figures} (Upper panel) $m_n^z$ for $N=100$ and different values of $\phi$; if $\phi$ deviates too much from $\pi$ we see that the oscillations of period $2\tau$ die away. (Lower panel) Corresponding power spectrum of the Fourier transform of the $z$-magnetization: when there is period doubling, two peaks around $\omega_B=\pi/\tau$ appear.\\
{\em Insets} (Upper panel) Dependence on $N$ of the main peaks frequencies $\omega_{\pm}(\phi,N)$ in the Fourier power spectrum: whenever there is the time crystal, these frequencies tend to $\pi/\tau$. (Lower panel) Dependence on $1/N$ of the main peaks height $|m_{\omega_{\pm}(\phi,N)}^z|^2$ in the Fourier power spectrum: whenever there is the time crystal, this height tends to a constant for $N\to\infty$, otherwise it tends to 0 as a power law (notice the logarithmic scale on both axes).
(Numerical parameters $h = 0.32,\,\tau=0.6$. 
The Fourier transforms in the insets are performed over $K=65536$ driving periods, in the main figures over $K=32768$ periods.)} 
  \label{fig2}
\end{figure}

The upper panel of Fig.~\ref{fig2} shows that $m_n^z$ displays period-doubling oscillations when the deviation of $\phi$ from $\pi$ is small 
enough (here $h_i=h$, $N=100$ and we consider different values of $\phi$). In the lower panel there are the plots of the corresponding Fourier power spectra $|m_\omega^z|^2$
vs $\omega$. Whenever there are persisting oscillations in the time domain, the power spectrum $|m_\omega^z|^2$ shows two peaks at frequencies $\omega_{\pm}(\phi,N)$ symmetric around $\omega_B$. The peaks are responsible for the oscillations with period $2\tau$ in $m_n^z$ with superimposed beatings occurring with a period $1/|\omega_{+}(\phi,N)-\omega_{-}(\phi,N)|$.

In order to convince us that the time-crystal behaviour is persistent in the thermodynamic limit (condition {\em III)}) and that the beatings disappear in this limit, we have to study the behaviour of the peaks at $\omega_{\pm}(\phi,N)$ when $N$ is increased. In the inset of the upper panel we show the dependence on $N$ of the frequencies $\omega_{\pm}(\phi,N)$ of the peaks, while in the inset of the lower panel we show the dependence on $1/N$ of the height $|m_{\omega_{\pm}(\phi,N)}^z|^2$ of the peaks (notice the logarithmic scale on the two axes). When the $2\tau$ oscillations die away ($\phi=0.84\pi$), we see that $\omega_{\pm}(\phi,N)$ tend to a limit {\em different} from $\omega_B$ when $N\to\infty$ and the height of the peaks $|m_{\omega_{\pm}(\phi,N)}^z|^2$ tends to 0 as a power law: in this case there is no time crystal. On the opposite, for all those values of $\phi$ for which we phenomenologically see persisting oscillations of period $2\tau$, $\omega_{\pm}(\phi,N)$ 
approach $\pi/\tau$ exponentially fast in $N$.~\cite{Note1} Moreover, for the same values of $\phi$, $|m_{\omega_{\pm}(\phi,N)}^z|^2$ tends to a constant for $N\to\infty$ (inset of the lower panel of Fig.~\ref{fig2}). We find indeed that there are persisting oscillations at $\omega_B$ and in this case there is a time crystal.

\begin{figure}
  \begin{tabular}{c}
   \hspace{0cm}\resizebox{80mm}{!}{\includegraphics{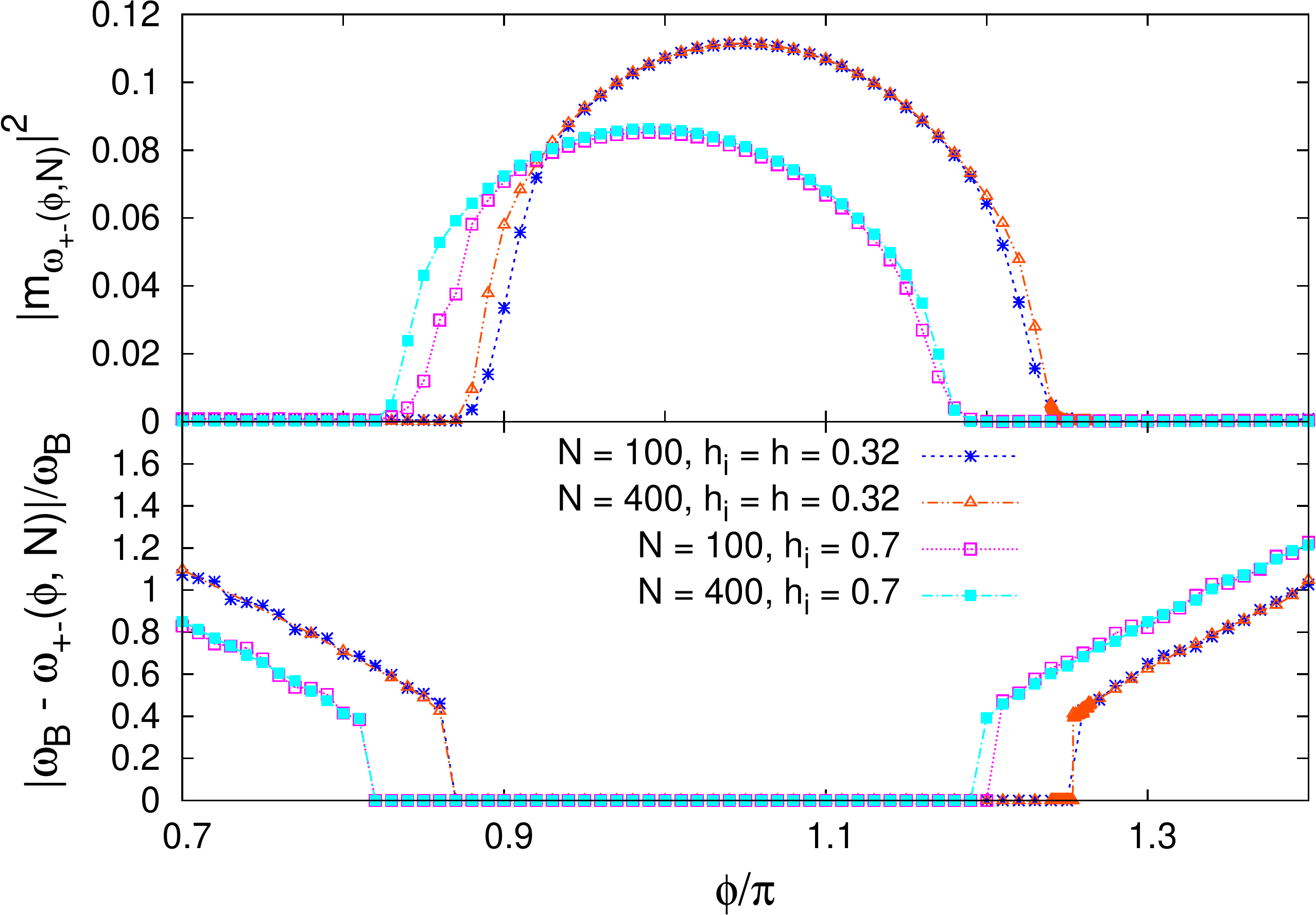}}\\
  \end{tabular}
\caption{(Upper panel) Dependence on $\phi$ of the main peaks height $|m_{\omega_{\pm}(\phi,N)}^z|^2$ for different $N$ and initial conditions: we see a full region around $\phi=\pi$ where it is different from 0. (Lower panel) Dependence on $\phi$ of the main peaks frequencies: when $|m_{\omega_{\pm}(\phi,N)}^z|^2$ is different from 0 the frequencies equal $\omega_B$ up to terms exponentially small in $N$ and the system is a time crystal. (Numerical parameters $h=0.32,\,\tau=0.6$, the Fourier transforms have been performed over $K=65536$ driving periods).}
  \label{figA}
\end{figure}

In the upper panel of Fig.~\ref{figA} we show how the main peaks height $|m_{\omega_{\pm}(\phi,N)}^z|^2$ depends on $\phi$ for different values of $N$ and of $h_{\rm i}$. We see a quite large region around $\phi=\pi$ where the peaks height is different from 0: in all this region the system behaves as a time crystal because the peak frequencies $\omega_{\pm}(\phi,N)$ equal $\omega_B$ up to terms exponentially small in $N$ (lower panel of the same figure). We see that $|m_{\omega_{\pm}(\phi,N)}^z|^2$ vanishes in a continuous way at the boundaries, while the frequencies jump in a discontinuous way. We see moreover that the region where the time-crystal behaviour occurs depends on the value of $h_{\rm i}$. 
{Moving outside of this region, the peaks height suddenly drops by two orders of magnitude.} 

In Section~\ref{classical:sec} we are going to show that the time-translation symmetry breaking transition in $\phi$ can be described in terms of the properties of the phase space of the effective classical Hamiltonian which describes the LMG model in the $N\to\infty$ limit.

%
%

\subsection{Robustness of the time-crystal phase: perturbations in the kicking operator}
\begin{figure}[h!]
  \begin{center}
    \begin{tabular}{cc}
      \hspace{-0.5cm}\resizebox{80mm}{!}{\includegraphics{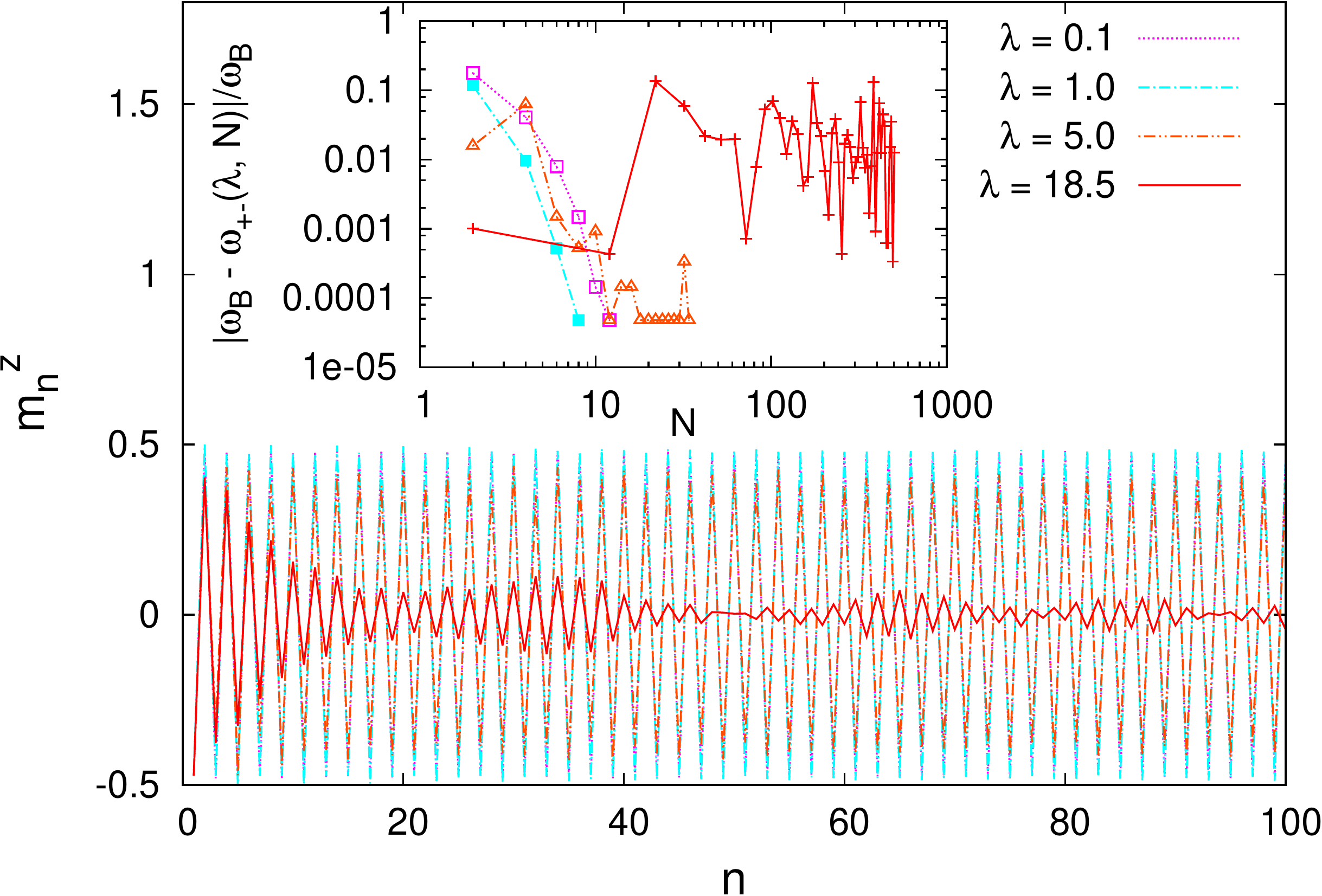}}\\
      \hspace{-0.5cm}\resizebox{80mm}{!}{\includegraphics{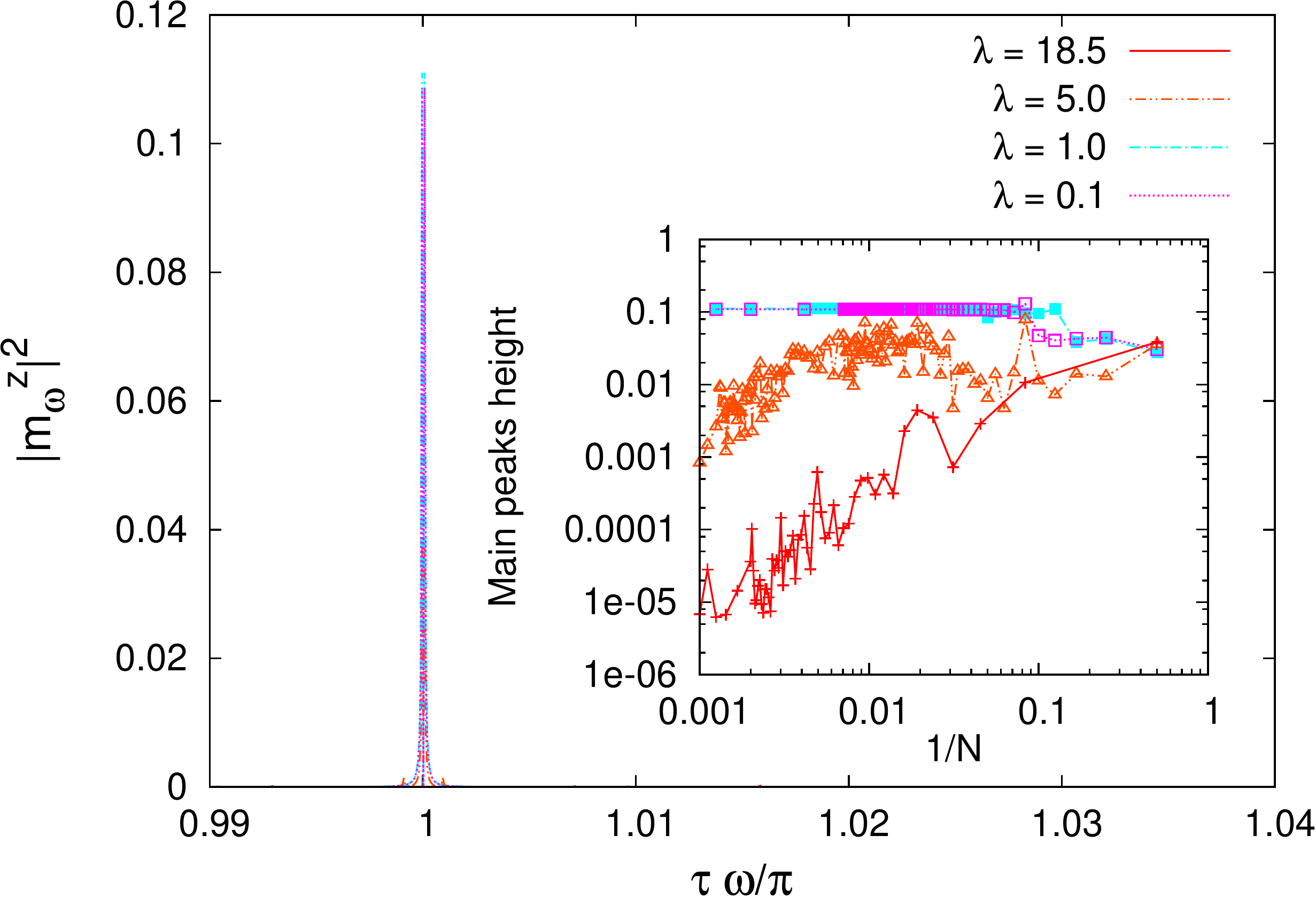}}\\
    \end{tabular}
  \end{center}
\caption{{\em Main figures} (Upper panel) Evolution of $m_n^z$ for $N=100$ and different values of $\lambda$. (Lower panel) Power spectra $|m_\omega^z|^2$ of the corresponding Fourier transforms.  {\em Insets} (Upper panel) Dependence on $N$ of the frequencies $\omega_{\pm}(\lambda,N)$ of the main peaks  in the power spectrum. (Lower panel) Height $|m_{\omega_{\pm(\lambda,N)}}^z|^2$ of the main peaks versus $1/N$ (double logarithmic scale). (Numerical parameters: $h_{\rm i} = h = 0.32,\, \tau = 0.6$. The Fourier transforms are performed over $K=65536$ driving periods in the insets and $K=32768$ periods in the main figures.)}
\label{lambda:fig}
\end{figure}

We can probe the rigidity of the period doubling also perturbing our kicking in a radical way: we choose a kicking capable to induce a quantum chaotic behaviour of the system~\cite{Haake_ZPB86,Khripkov_PRE13}. Inspired by Ref.~\onlinecite{Haake_ZPB86} 
we choose a time-evolution operator over one period of the form
\begin{equation} \label{eckickazzone:eqn}
  \widehat{U}_{\lambda}=\exp\left[-i\phi \opS^{\,x}\right]\exp\left[i\lambda(\opS^z)^2/N\right]\exp\left[-i\tau\widehat{H}(h)\right]\,.
\end{equation}
We fix $\phi=\pi$, fix also $\tau$ and $h$, and we consider different values of $\lambda$. Results are reported in Fig.~\ref{lambda:fig}. On the top panel we show the evolution of $m_n^z$: we phenomenologically see that there are persistent oscillations for $\lambda$ small while there are collapses and revivals for larger $\lambda$. Looking at the power spectrum of the Fourier transform of $m_n^z$ (lower panel) we see two peaks at $\omega_{\pm}(\lambda,N)$ around $\omega_B$ which, for $\lambda$ large, become very small. Nevertheless, some discernible features around $\omega_B$ still persist. 

As before, in order to inquire the persistence of the time-crystal behaviour in the thermodynamic limit, we study how those peaks depend on $N$. In the inset of the upper panel of Fig.~\ref{lambda:fig} we show their frequencies: we see that they tend to $\omega_B$ only for some values $\lambda$.~\cite{Note1} 
Correspondingly, the height of those peaks tends to a non-vanishing value for $N\to\infty$ only for some values of $\lambda$; for others it tends to 0 as a power law (lower panel of Fig.~\ref{lambda:fig}). We have a time crystal only when, in the limit $N\to\infty$, there is a finite-amplitude response at frequency $\omega_B$. Among the cases we consider, only $\lambda=0.1$ and $\lambda=1.0$ meet this condition. In the case $\lambda=5.0$ we have $\omega_{\pm}(\lambda,N)\to\omega_B$ for $N\to\infty$ but the height of the peak goes to zero; when $\lambda=18.5$, instead, none of the two conditions is met: the height goes to zero and the frequency does not tend to $\pi/\tau$. In Section~\ref{classical:sec} we better discuss how the time-crystal behaviour depends on $\lambda$, also in connection with chaotic properties of the model. In that section we also show how the presence of the time-translation symmetry breaking depends on $\lambda$ (see the left panel of Fig.~\ref{figklass}).

\subsection{Floquet states} 
\label{Nayak:sec}
\begin{figure}[h!]
  \begin{center} 
    \begin{tabular}{cc}
      \hspace{-0.cm}\resizebox{80mm}{!}{\includegraphics{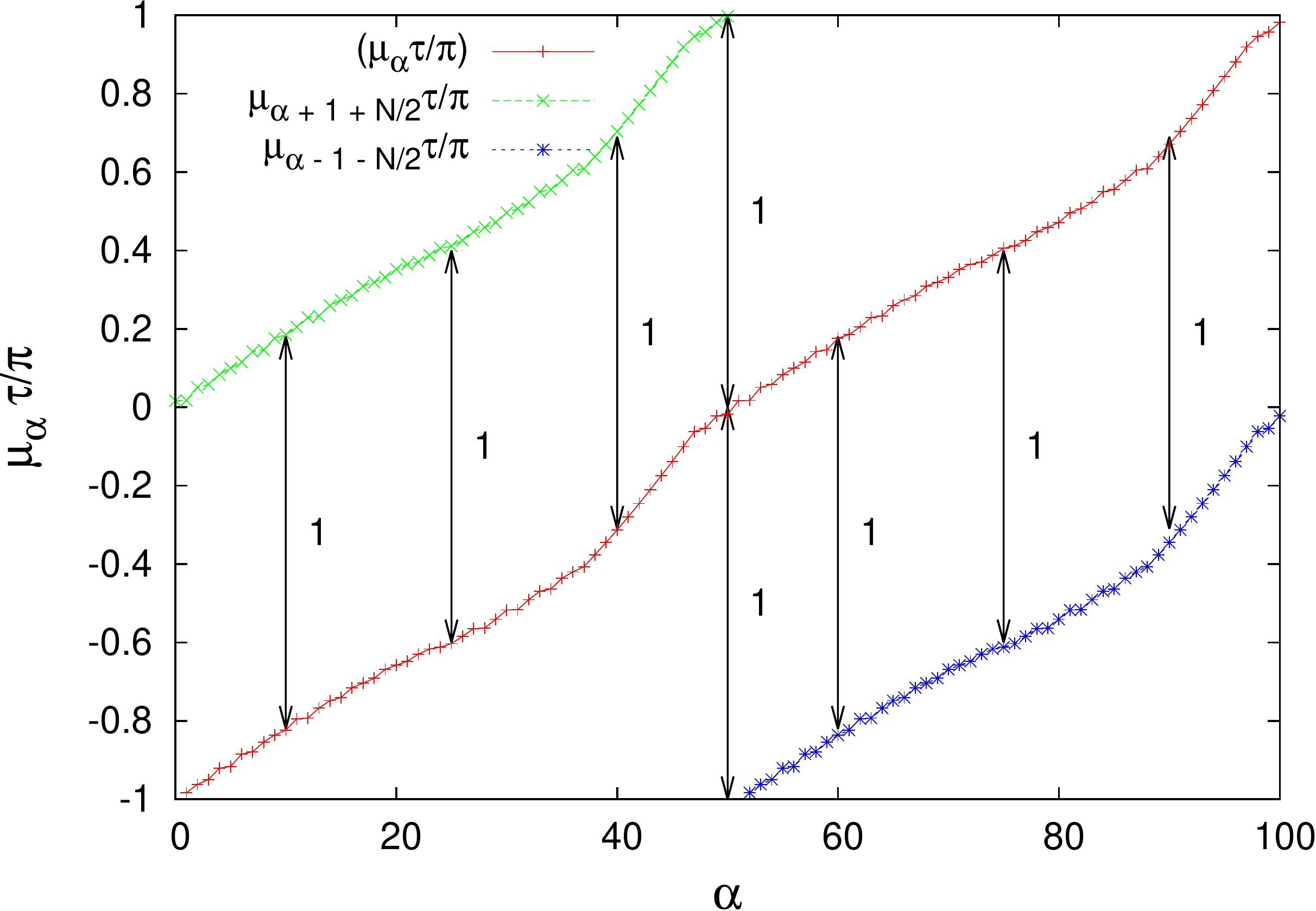}}\\
      \hspace{0.cm}\resizebox{80mm}{!}{\includegraphics{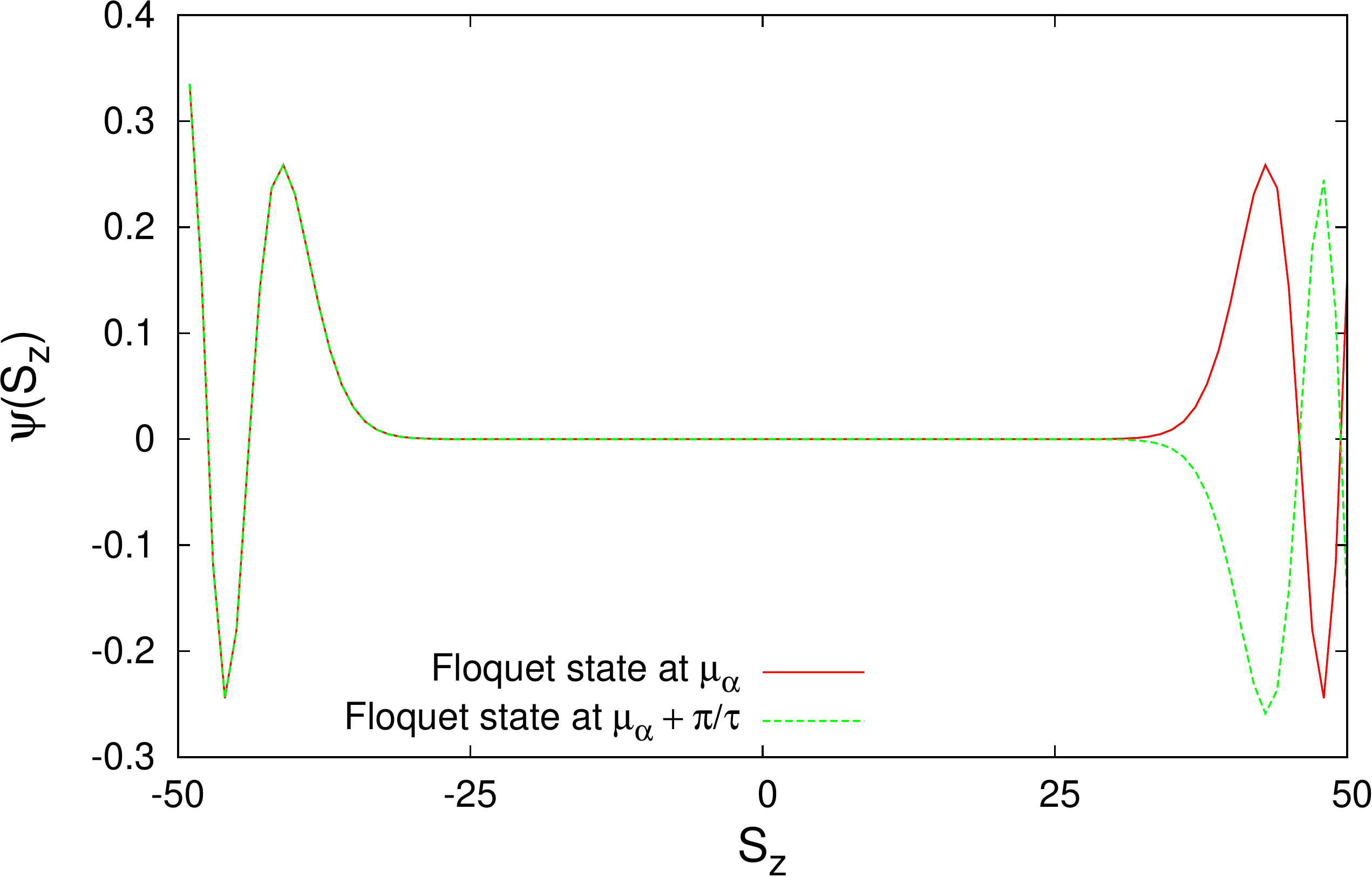}}\\
    \end{tabular}
  \end{center}
\caption{\red{(Upper panel) {Three copies of the Floquet spectrum vs $\alpha$, horizontally shifted with respect to each other by $1+N/2$ and divided by $\pi/\tau$. 
The three curves constantly differ from each other by 1 along the vertical axis: each quasi-energy has its own partner shifted by $\pm\pi/\tau$.} 
(Lower panel) The corresponding Floquet states are organised in pairs: in each pair the quasi-energies differ by $\pi/\tau$ and the two states are even and odd superpositions of symmetry broken states (we plot the (real) amplitudes of the members of one of such pairs in the basis of the eigenstates of $\widehat{S}^z$). Numerical parameters: $N=100,\,\tau=0.006,\,h_0=0.32,\,\phi=\pi,\,\lambda=0$.} }
\label{verifica}
\end{figure}

{Refs.~\onlinecite{Nayak_PRL16,Vedika_PRL16,Keyser_PRB16} state} that in a time crystal the eigenstates of the dynamics need to have very specific properties. In order to understand these properties, we analyse these eigenstates in our periodically driven setting. They are the Floquet states $\ket{\phi_\alpha}$, defined as the eigenstates of the time-evolution operator over one period; the phases of the corresponding eigenvalues are the quasi-energies $\mu_\alpha$~\cite{Shirley_PR65} 
\begin{equation}
  \widehat{U}(\tau,0)=\sum_\alpha\nep^{-i\mu_\alpha\tau}\ket{\phi_\alpha}\bra{\phi_\alpha}\,.
\end{equation}
We can see that the Floquet states are eigenstates of the stroboscopic dynamics and, after a time $n\tau$, they acquire a phase factor $\nep^{-in\mu_\alpha\tau}$. Given these definitions, the authors of {Refs.~\onlinecite{Nayak_PRL16,Vedika_PRL16,Keyser_PRB16}} find that, in order to obtain a time crystal with period doubling, one needs to have a Floquet spectrum with a specific structure. In particular any Floquet state with quasi-energy $\mu_\alpha$ needs to have a partner with a quasi-energy $\mu_\alpha+\pi/\tau$. Each of these pairs behaves as the even and the odd superposition of two symmetry-broken states. Preparing the system in a symmetry broken state, it undergoes Rabi oscillations between these two pairs with a frequency $\pi/\tau$, the difference of the eigenfrequencies of these two states. It is crucial that all the Floquet spectrum, or at least an extensive fraction of it, shows this doublet structure in order to see the time-translation symmetry breaking in the observables~\cite{Nayak_PRL16,Emanuele_arXiv16}. We are indeed able to directly check that our system obeys these properties in the special case of $\phi=\pi$ and $\tau$ small (in order to not have spectrum folding). In this case we can check that each quasi-energy has its partner shifted by an amount $\pi/\tau$. \red{We see this fact in the upper panel of Fig.~\ref{verifica}: here we plot {three copies} of the same spectrum vs $\alpha$, {horizontally shifted with respect to each other by $1+N/2$ and divided by $\pi/\tau$. We see that the three curves constantly differ by 1 along the vertical axis: for each quasi-energy $\mu_\alpha$ there is a partner shifted by $\pm\pi/\tau$ (the sign is not important being the Floquet spectrum periodic by $2\pi/\tau$). }
\red{The same periodicity in the Floquet spectrum for an LMG model, with a different form of driving, has been found in Ref.~\onlinecite{PRA_Holthaus}.} 
In the lower panel of Fig.~\ref{verifica} we show two Floquet states whose quasi-energies differ by $\pi/\tau$ ($\pi$-paired states). We see 
that these two $\pi$-paired states are respectively even and odd superpositions of $\mathbb{Z}_2$-symmetry breaking states: the situation is therefore strictly analogous to the one found in Refs.~\onlinecite{Nayak_PRL16,Vedika_PRL16,Keyser_PRB16}, which we have reviewed above.}

\begin{figure}[h!]
  \begin{center} 
    \begin{tabular}{cc}
      (a)\\
      \hspace{-0.cm}\resizebox{80mm}{!}{\includegraphics{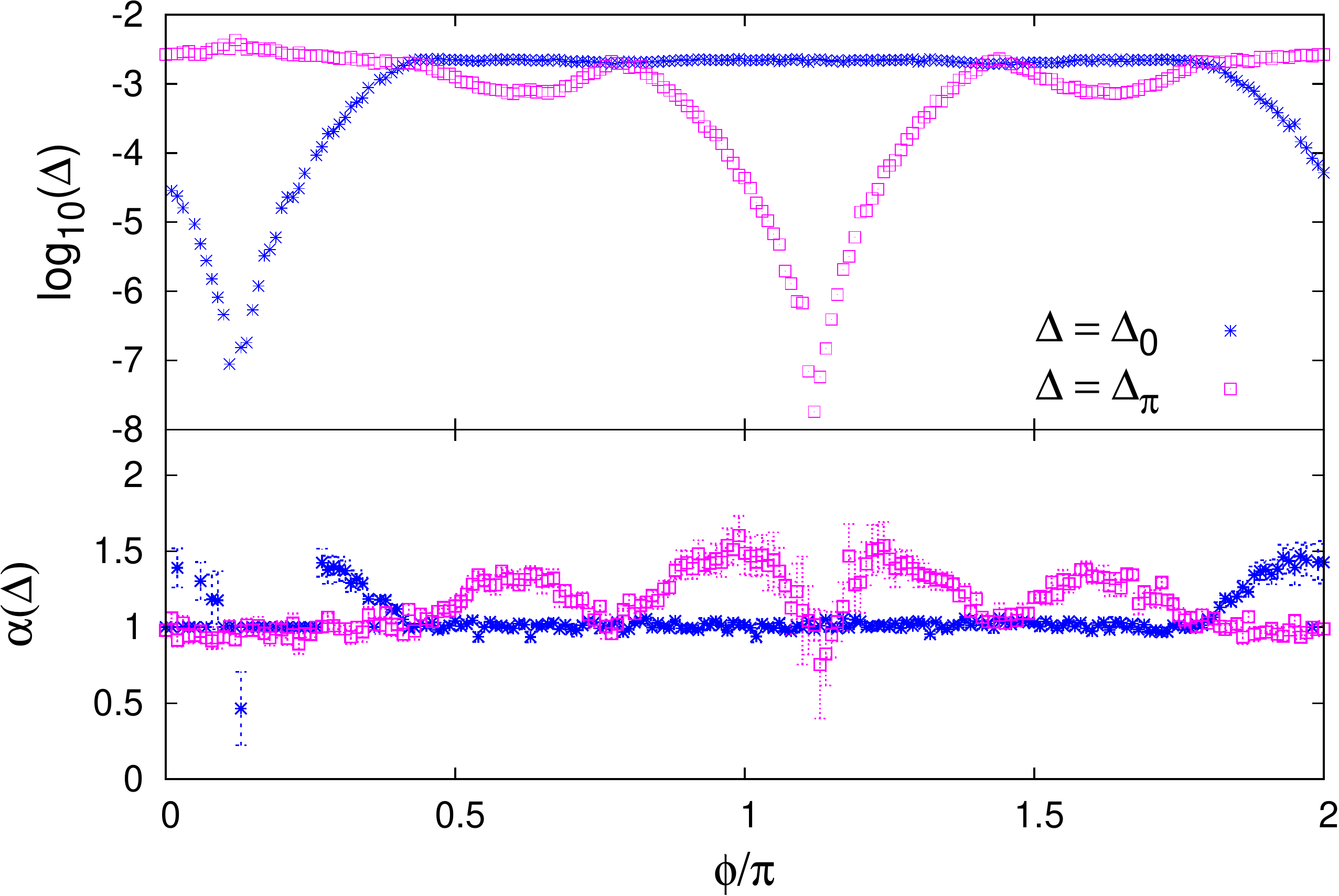}}\\
      (b)\\
      \hspace{0.cm}\resizebox{80mm}{!}{\includegraphics{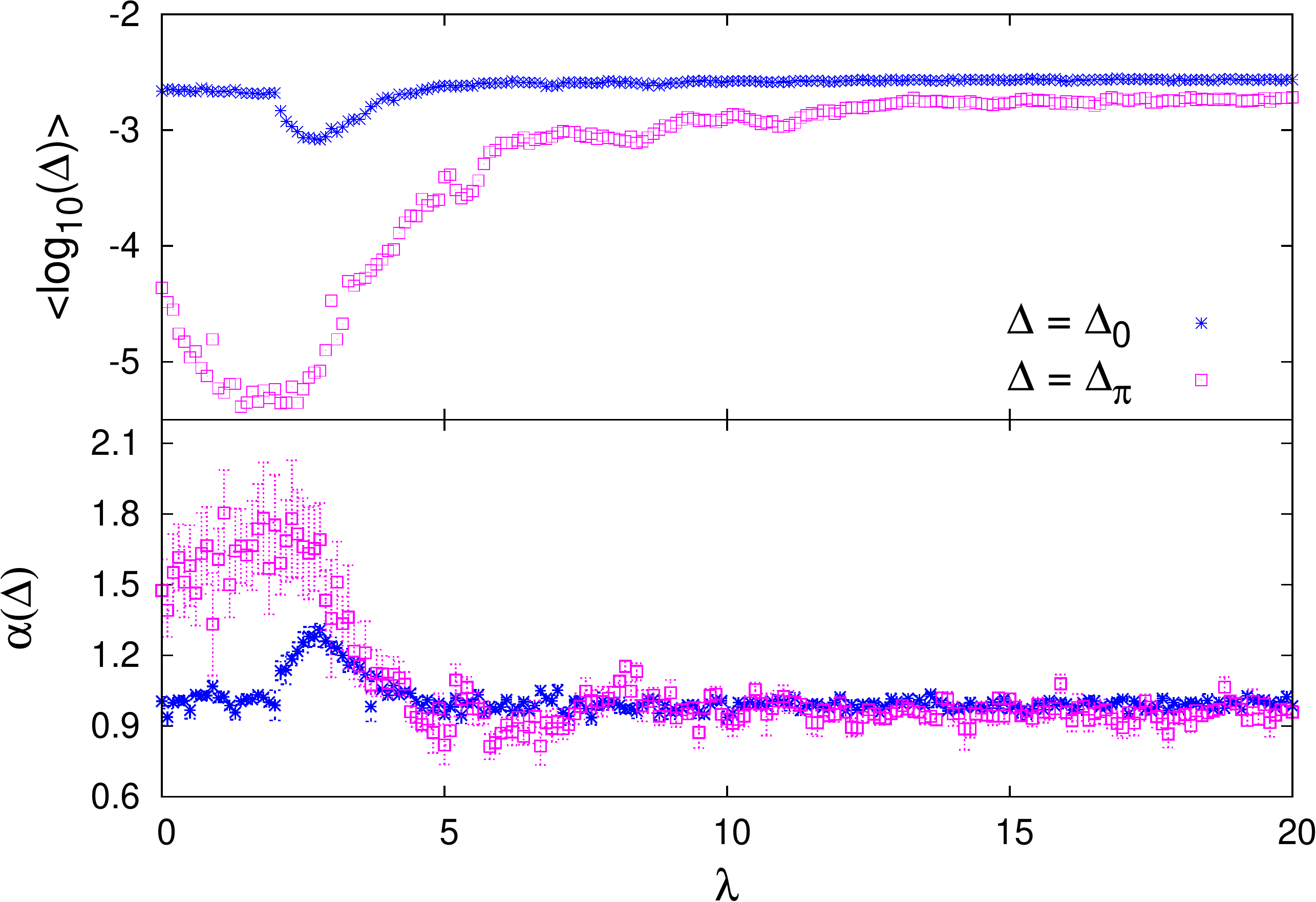}}\\
    \end{tabular}
  \end{center}
\caption{\red{Panel (a): (upper plot) plot of $\mean{\log_{10}\Delta_0}$ and $\mean{\log_{10}\Delta_\pi}$ (see Eq.~\eqref{logelt:eqn}) vs $\phi$ fixing $\lambda=0$; (lower plot) plot of the corresponding fitting exponents $\alpha(\Delta_0)$ and $\alpha(\Delta_\pi)$ vs $\phi$ (see Eq.~\eqref{fittola:eqn}). Panel (b): (upper plot) plot of $\mean{\log_{10}\Delta_0}$ and $\mean{\log_{10}\Delta_\pi}$ (see Eq.~\eqref{logelt:eqn}) 
vs $\lambda$ fixing $\phi=\pi$; (lower plot) plot of the corresponding fitting exponents $\alpha(\Delta_0)$ and $\alpha(\Delta_\pi)$ vs $\lambda$ (see Eq.~\eqref{fittola:eqn}).
As expected, there is $\pi$-spectral pairing -- $10^{\mean{\log_{10}\Delta_0}}\ll10^{\mean{\log_{10}\Delta_\pi}}$ and $\alpha(\Delta_\pi)>\alpha(\Delta_0)$-- when there is time-crystal behaviour. (Numerical parameters: $\tau=0.6,\,h=0.32,\,N=1600$ in the upper plots.) } }
\label{verifica1}
\end{figure}

\red{The direct check of the $\pi/\tau$ periodicity of the spectrum (or at least of an extensive part of it) is possible only for $\tau$ very small. The reason is that, being the quasienergies obtained as phases, they are defined up to translations of $2\pi/\tau$. In particular, it is possible to fold all the spectrum in the so-called first Brillouin zone $[-\pi/\tau,\pi/\tau]$: the numerical algorithms evaluate the quasienergies folded in this interval. 
If $\tau$ is small enough, the bandwidth of the Floquet spectrum is smaller than $2\pi/\tau$ and there is no folding in the numerically evaluated quasi-energies. If instead there is folding, as in the case $\tau=0.6$ that we consider throughout the paper, we need a different strategy to check if an extensive part of the spectrum is organized in pairs shifted by $\pi/\tau$ ($\pi$-spectral pairing) in the thermodynamic limit. As done in Ref.~\onlinecite{Vedika_PRB16}, we consider the level spacings $\Delta_0^{(\alpha)}=\mu_{\alpha+1}-\mu_\alpha>0$ and the $\pi$-translated level spacings 
$$\Delta_\pi^{(\alpha)}=\min_{\beta\,|\, \mu_\beta> (\mu_\alpha+\pi/\tau)_1}\left[\mu_\beta-(\mu_\alpha+\pi/\tau)_1\right]>0\,.$$ 
The symbol $(\cdots)_1$ means that we translate the argument by a multiple of $2\pi/\tau$, so that it falls inside the first Brillouin zone. Now we perform the averages 
\begin{eqnarray} \label{logelt:eqn}
  \mean{\log_{10}\Delta_0}&=&\frac{1}{N+1}\sum_{\alpha=1}^{N+1}\log_{10}\Delta_0^{(\alpha)}\nonumber\\
 \mean{\log_{10}\Delta_\pi}&=&\frac{1}{N+1}\sum_{\alpha=1}^{N+1}\log_{10}\Delta_\pi^{(\alpha)}\,.
\end{eqnarray}
 If the system is a time crystal and there is $\pi$-spectral pairing in the thermodynamic limit, we expect $10^{\mean{\log_{10}\Delta_\pi}}$ to be much smaller than $10^{\mean{\log_{10}\Delta_0}}$ even for $N$ finite.~\cite{Vedika_PRB16} Moreover, in order to have $\pi$-spectral pairing in the thermodynamic limit, we need that $10^{\mean{\log_{10}\Delta_\pi}}$ scales to 0 faster than $10^{\mean{\log_{10}\Delta_0}}$ for $N\to\infty$,~\cite{Vedika_PRB16} namely
\begin{equation} \label{nullezza:eqn}
  \lim_{N\to\infty}10^{\mean{\log_{10}\Delta_\pi}-\mean{\log_{10}\Delta_0}}=0\,.
\end{equation}
From the numerical results, in this model we empirically verify the relations 
\begin{equation} \label{fittola:eqn}
 \mean{\log_{10}\Delta_{0/\pi}}\sim\beta(\Delta_{0/\pi})-\alpha(\Delta_{0/\pi})\log_{10}N\,
\end{equation}
where $\beta(\Delta_{0/\pi})$ and $\alpha(\Delta_{0/\pi})$ are numerically found coefficients. We have therefore that both the gaps scale with $N$ as a power law.  Therefore, in our system, checking Eq.~\eqref{nullezza:eqn} is equivalent to check that
$$
  \alpha_\pi > \alpha_0\,.
$$
Linearly fitting $\mean{\log_{10}\Delta_{0/\pi}}$ vs $\log_{10}N$ with the minimum square algorithm, we numerically obtain $\alpha(\Delta_0)$ and $\alpha(\Delta_\pi)$.
We show our results in Fig.~\ref{verifica1}. On the upper plot of panel (b) we plot $\mean{\log_{10}\Delta_\pi}$ and $\mean{\log_{10}\Delta_0}$ vs $\lambda$ for fixed $\phi=\pi$ and $N=1600$; on the lower plot the corresponding $\alpha_\pi$ and $\alpha_0$. We can see that, for approximately $\lambda<5$, $10^{\mean{\log_{10}\Delta_\pi}}$ is some orders of magnitude smaller than $10^{\mean{\log_{10}\Delta_0}}$ and $\alpha_\pi > \alpha_0$. Therefore, in this range of $\lambda$ there is $\pi$-spectral pairing: comparing with the right panel of Fig.~\ref{figklass} we see that this range of $\lambda$ corresponds to a clear time-crystal behaviour, confirming our expectations. In Fig.~\ref{verifica1}(a) we show the $\phi$-dependence of $\mean{\log_{10}\Delta_\pi}$ and $\mean{\log_{10}\Delta_0}$ for $N=1600$ (upper plot) and $\alpha_\pi$ and $\alpha_0$ (lower plot), fixing $\lambda=0$. We can see that $10^{\mean{\log_{10}\Delta_\pi}}$ is smaller than $10^{\mean{\log_{10}\Delta_0}}$ in some intervals around $\phi=\pi$; in these same intervals we see also $\alpha_\pi > \alpha_0$: here we find $\pi$-spectral pairing. The $\pi$-spectral pairing corresponds to time-translation symmetry breaking: 
it is around $\phi=\pi$ that the time-crystal behaviour appears (compare with Fig.~\ref{figA}).}

\section{Classical limit}
\label{classical:sec}
The transition from time-translation symmetry breaking to its absence can be better physically interpreted when $N$ is actually infinite and the system behaves classically: it is effectively described by the Hamiltonian~\cite{Bapst_JSTAT12,Sciolla_JSTAT11,Russomanno_EPL15}
\begin{equation} \label{classical_Ham:eqn}
  \calH(Q,P,t) = \calH_0(Q,P)+\calH_{\rm kick}(Q,P)\sum_n\delta(t-n\tau)\;,
\end{equation}
where
\begin{equation}
  \calH_0(Q,P)\equiv- \frac{1}{2}J Q^2 - {h} \; \sqrt{1-Q^2} \; \cos{(2P)} \;
\end{equation}
and
\begin{equation}
  \calH_{\rm kick}(Q,P) = - \frac{\lambda}{4} Q^2 + \frac{1}{2}\phi \; \sqrt{1-Q^2} \; \cos{(2P)} \;.
\end{equation}
(In this classical limit, the components of the magnetization depend on $Q$ and $P$ as $m^z=\frac{1}{2}Q$, $m^x=\frac{1}{2}\sqrt{1-Q^2}\cos(2P)$ and $m^y=\frac{1}{2}\sqrt{1-Q^2}\sin(2P)$). The $\mathbb{Z}_2$ symmetry of the quantum spin model is reflected in the symmetry $P\to -P$, $Q\to -Q$ of the classical limit Hamiltonian Eq.~\eqref{classical_Ham:eqn}.

Below we look at the Poincar\'e sections~\cite{Berry_regirr78:proceeding} of the dynamics of this Hamiltonian, with the same kinds of kicking considered
above. Constructing a 
Poincar\'e section is very simple: we take some initial values and we evolve them under the stroboscopic dynamics reporting on a 
$P$, $Q$ plot the sequence of the positions. If the initial condition is in a regular region of the phase space, our points will be over a 
one-dimensional manifold. If instead the initial condition is in a chaotic region of the phase space, our points will fill a two-dimensional portion 
of phase space. 

\begin{figure*}[h!]
  \begin{center} 
    \begin{tabular}{cc}
      \hspace{-1cm}\resizebox{80mm}{!}{\includegraphics{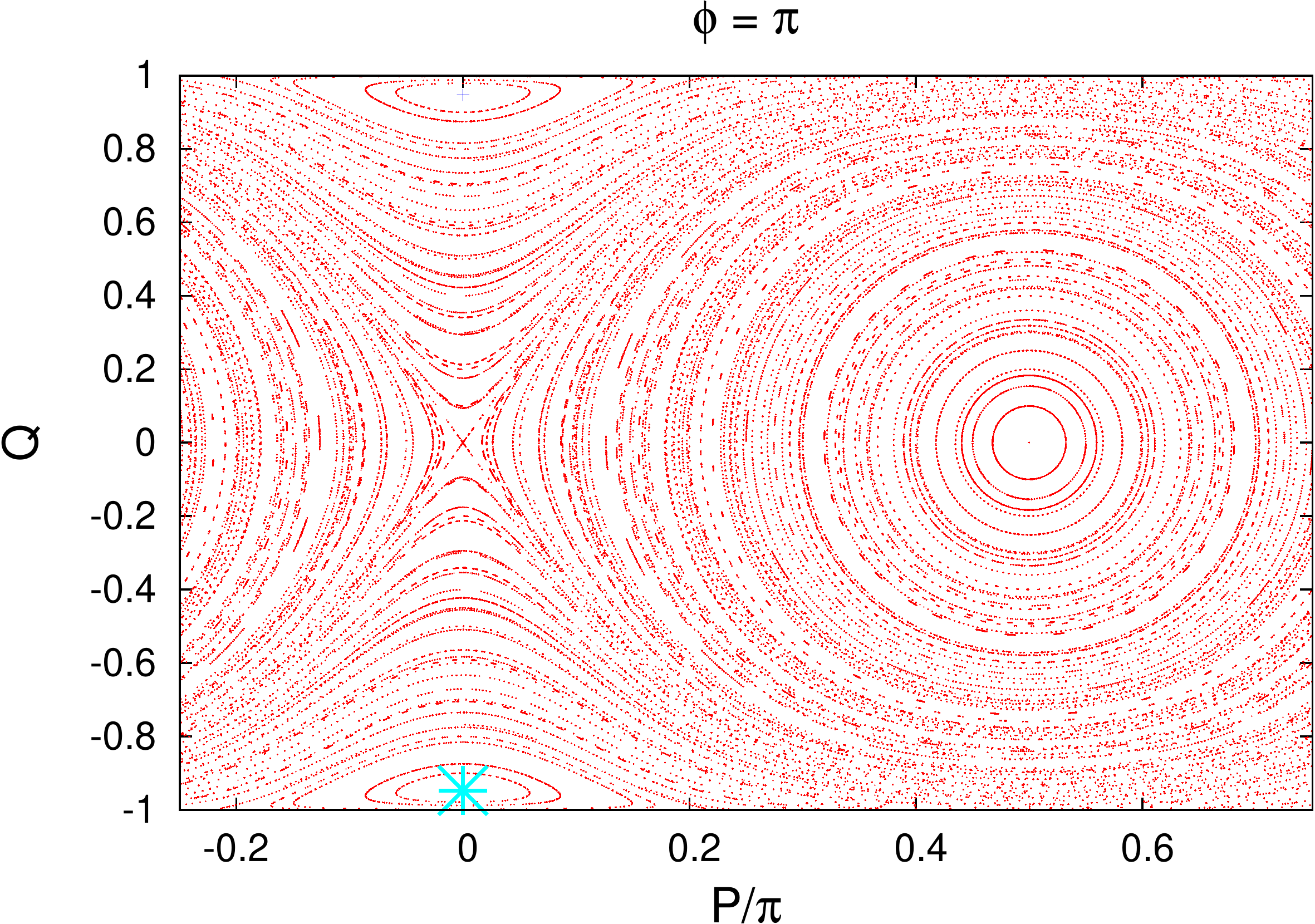}}&
      \hspace{0.4cm}\resizebox{80mm}{!}{\includegraphics{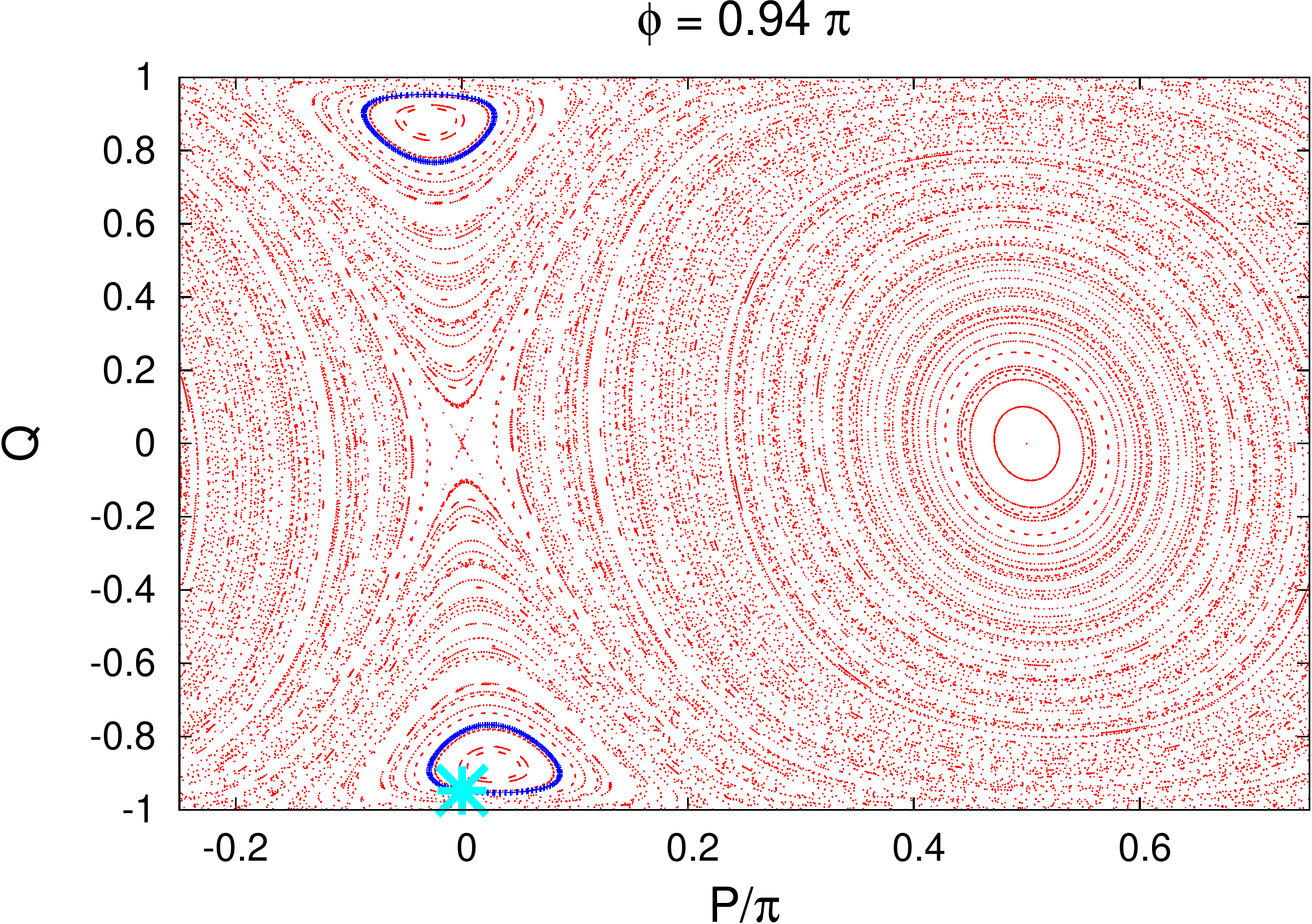}}\\
      \quad\\
      \hspace{-1cm}\resizebox{80mm}{!}{\includegraphics{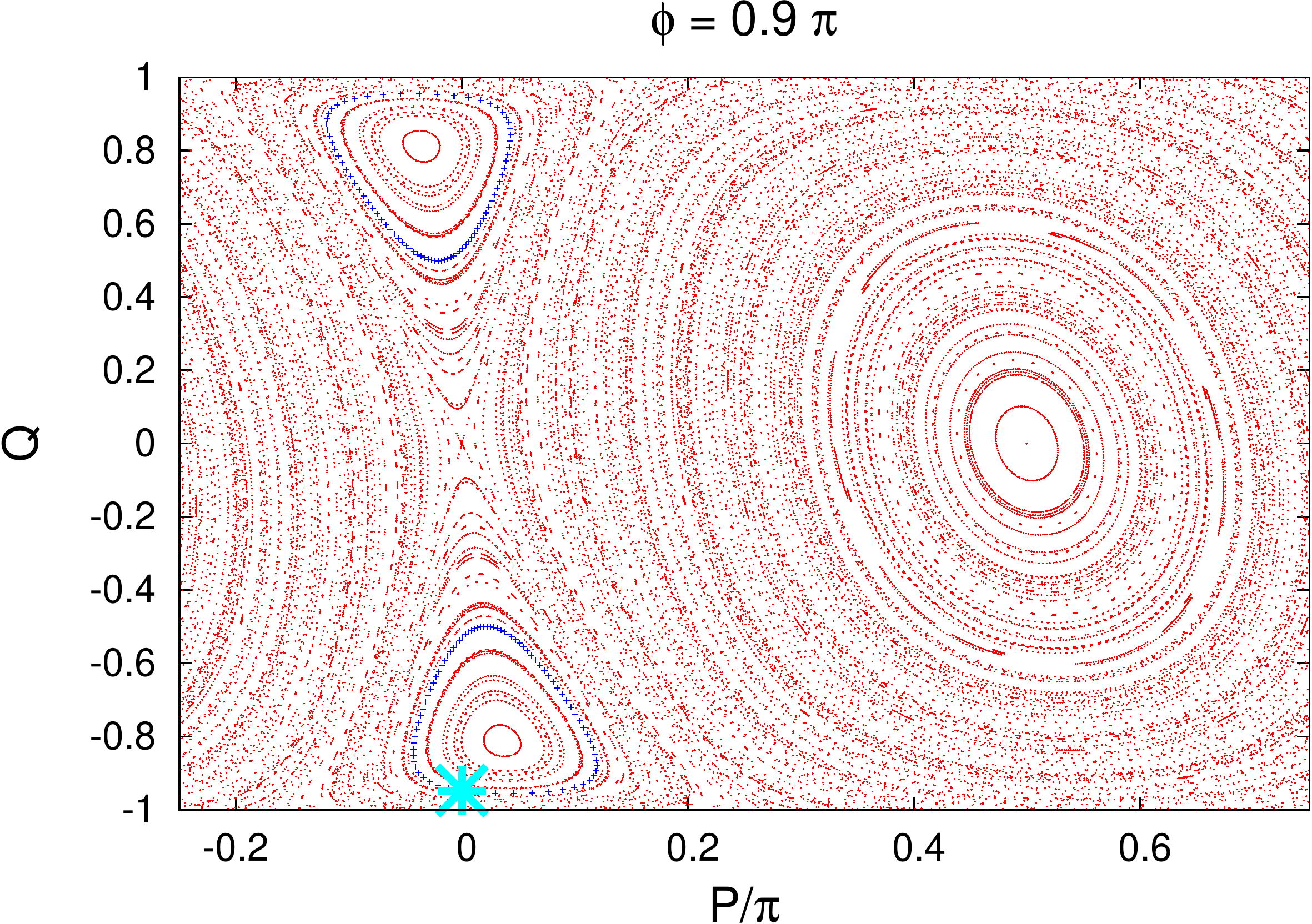}}&
      \hspace{0.4cm}\resizebox{80mm}{!}{\includegraphics{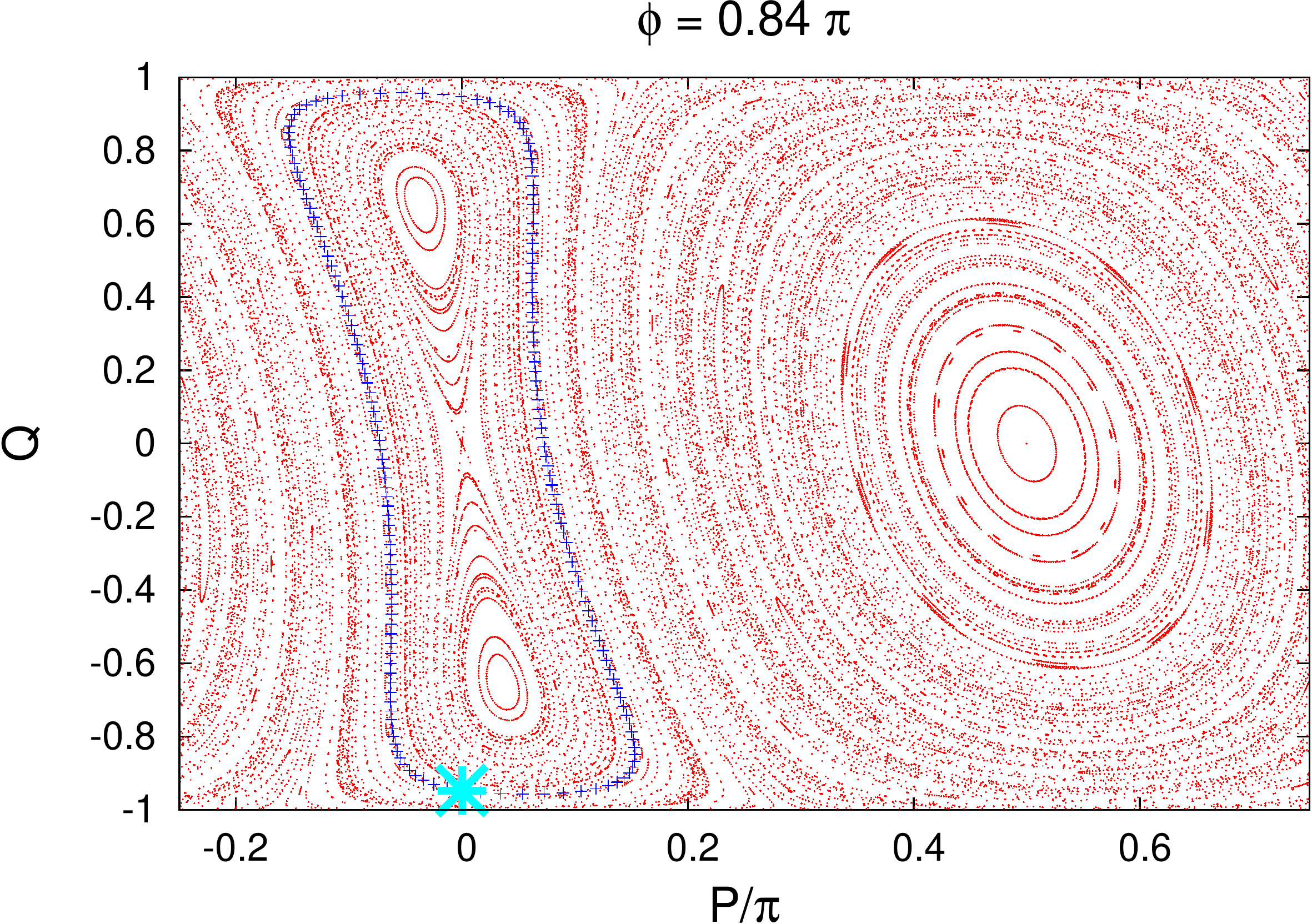}}\\
      \quad\\
      \hspace{-1cm}\resizebox{80mm}{!}{\includegraphics{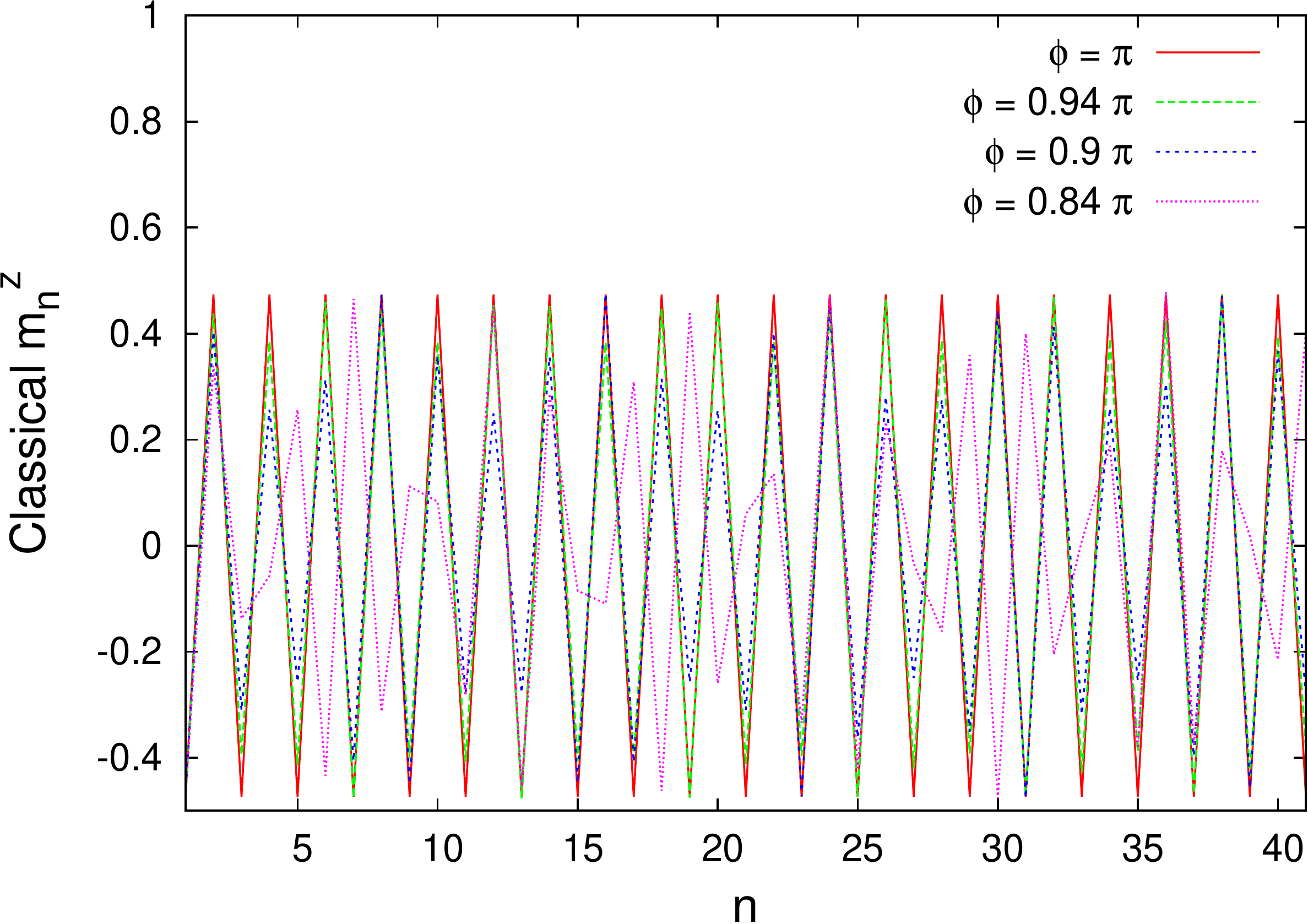}}&
      \hspace{0.4cm}\resizebox{80mm}{!}{\includegraphics{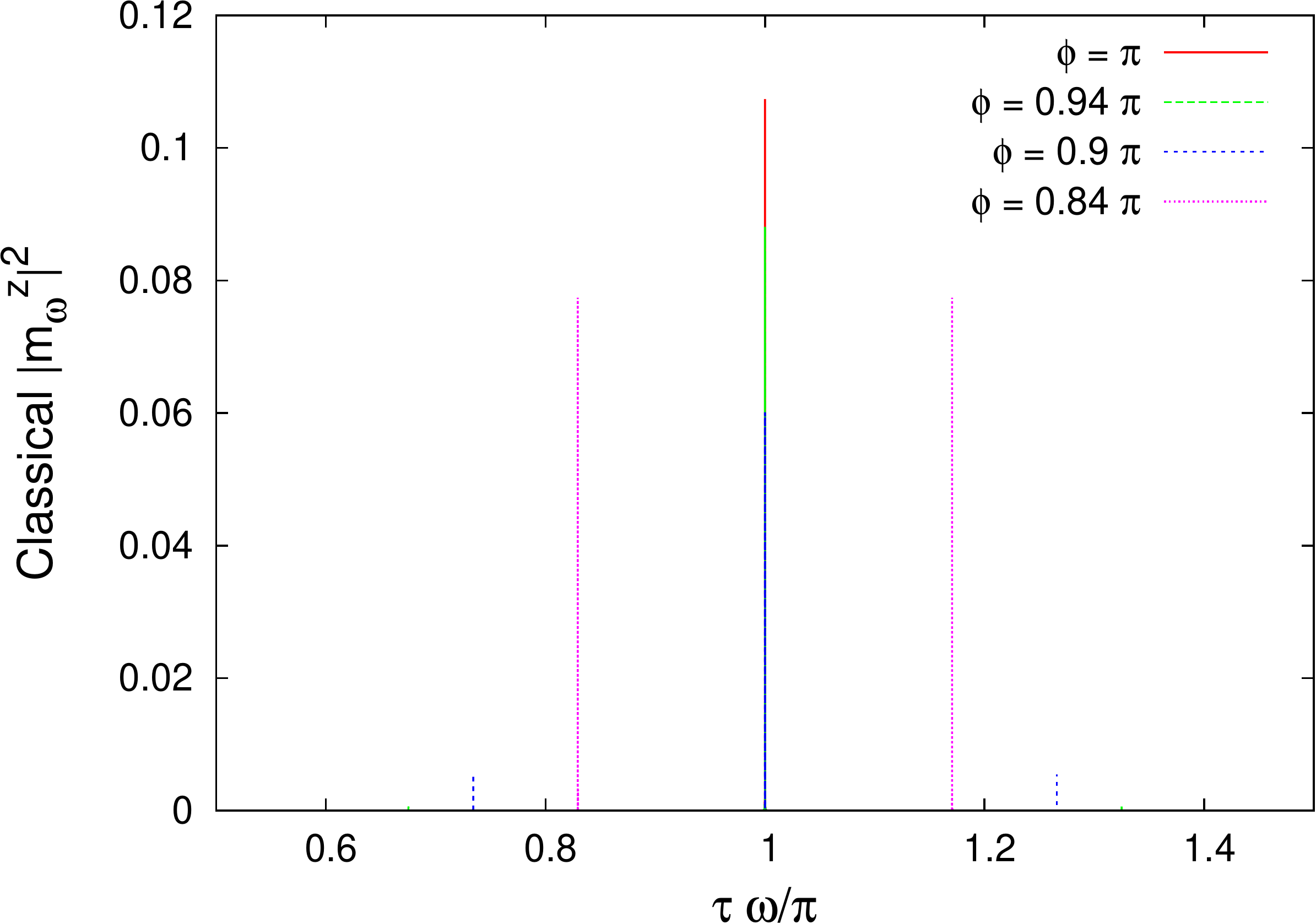}}
    \end{tabular}
  \end{center}
\caption{ (Upper and central panels) Classical phase space for different values of the phase $\phi$ and $\lambda=0$: the blue point corresponds to the initial symmetry-breaking ground state with $h_{\rm i}=h$. (Lower panel) Dynamics of the classical $m_n^z$ (left panel) and its Fourier transform (right panel): we initialize with the point in the classical phase space corresponding to the ground state for $h_{\rm i}=h$. (Numerical parameters: $h=0.32,\,\tau=0.6,\,\lambda=0$. The Fourier transform is performed over $K=65536$ driving periods.)}
\label{ponka_phi:fig}
\end{figure*}

In Fig.~\ref{ponka_phi:fig}  the different Poincar\'e sections are plotted, when $\lambda=0$, for different values of $\phi$: we see that the dynamics 
is always regular and each trajectory is a closed curve. We can moreover see that some curves are symmetric under the symmetry of the Hamiltonian $P\to -P$, $Q\to -Q$; while others break this symmetry. Each of the symmetry-breaking curves has a symmetric partner under $P\to -P$, $Q\to -Q$. The light-blue star in the graphs represents the initial condition in the classical limit, while the small blue crosses represent its stroboscopic evolution. We have put for clarity only the representative point for the initial symmetry-breaking 
ground state with $h_{\rm i}=h$; in the cases with $h_{\rm i}\neq h$ the argument runs exactly the same. 

On decreasing $\phi$ from $\pi$, we move from time-translation symmetry breaking to its absence (see Subsection~\ref{rigido_come_rocco:sec} and Fig.~\ref{figA}). In Fig.~\ref{ponka_phi:fig} we consider 3 values of $\phi$ for which there is time-translation symmetry breaking and one for which there is not ($\phi=0.84\pi$). In the first three cases the representative point of the initial 
state is on a curve breaking the symmetry $P\to -P$, $Q\to -Q$: it is trapped on this curve until the kick shifts it to the symmetric one with opposite sign of $Q$ (the two symmetric curves are highlighted in blue in the figures). 
In this way the sign of $m^z=\frac{1}{2}Q$ changes at any kick and the time-crystal behaviour arises. In the case $\phi=0.84\pi$, on the opposite, the representative point is 
on a curve invariant under the $P\to -P$, $Q\to -Q$
symmetry (highlighted in blue in the figure): there is no time crystal (central right panel of Fig.~\ref{ponka_phi:fig}).  The situation for $h\neq h_{\rm i}$ is very similar, the only difference is that the initial point is in a different position in the phase space, so 
it moves from a symmetry-breaking curve to a symmetric one at a different value of $\phi$. \red{The existence of $2\tau$-oscillations in a driven Hamiltonian similar to Eq.~\eqref{classical_Ham:eqn} and their connection with the Poincar\'e section properties have been also discussed in Ref.~\onlinecite{PRA_Holthaus}.}

\begin{figure*}
  \begin{tabular}{cc}
   \resizebox{80mm}{!}{\includegraphics{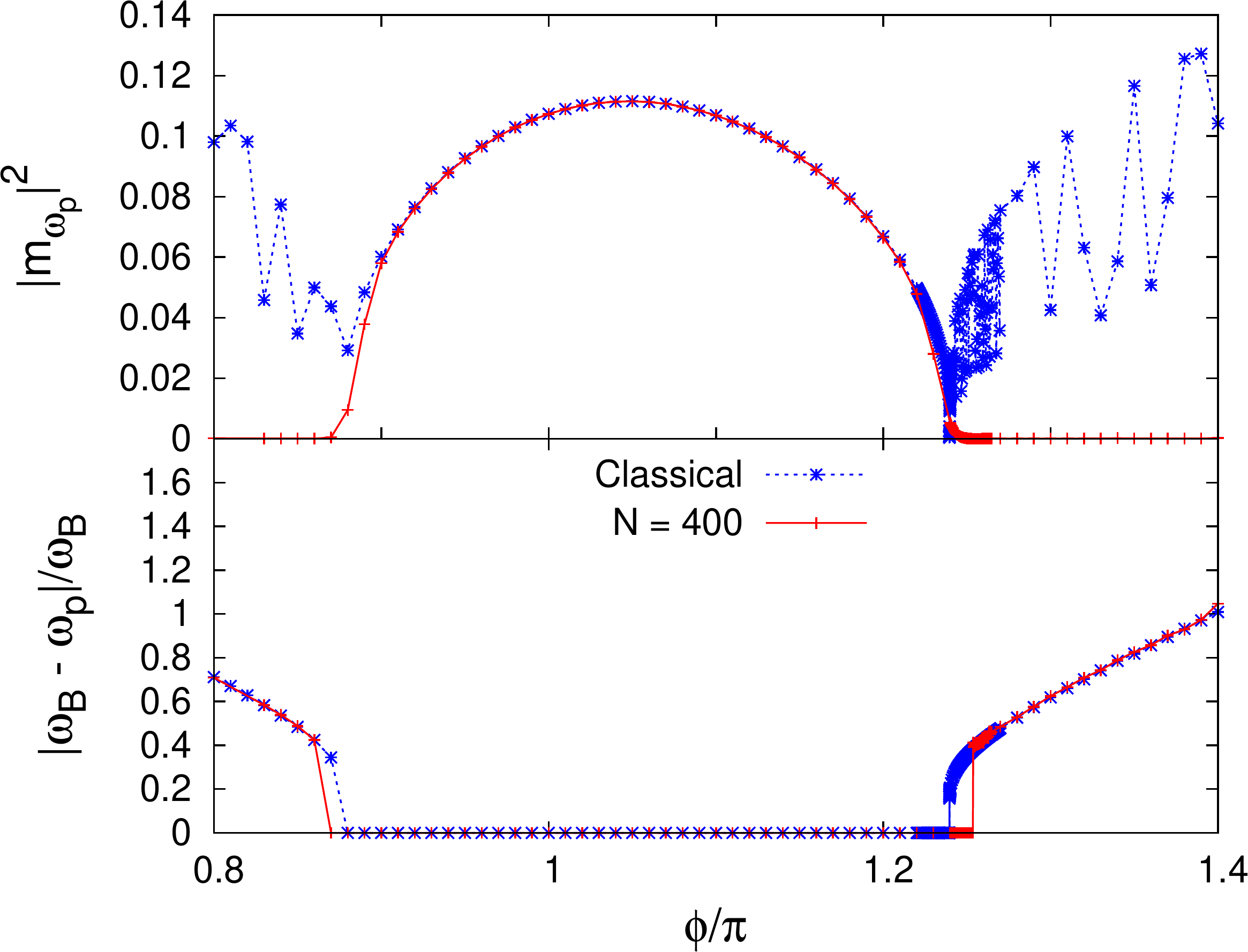}}&
   \resizebox{80mm}{!}{\includegraphics{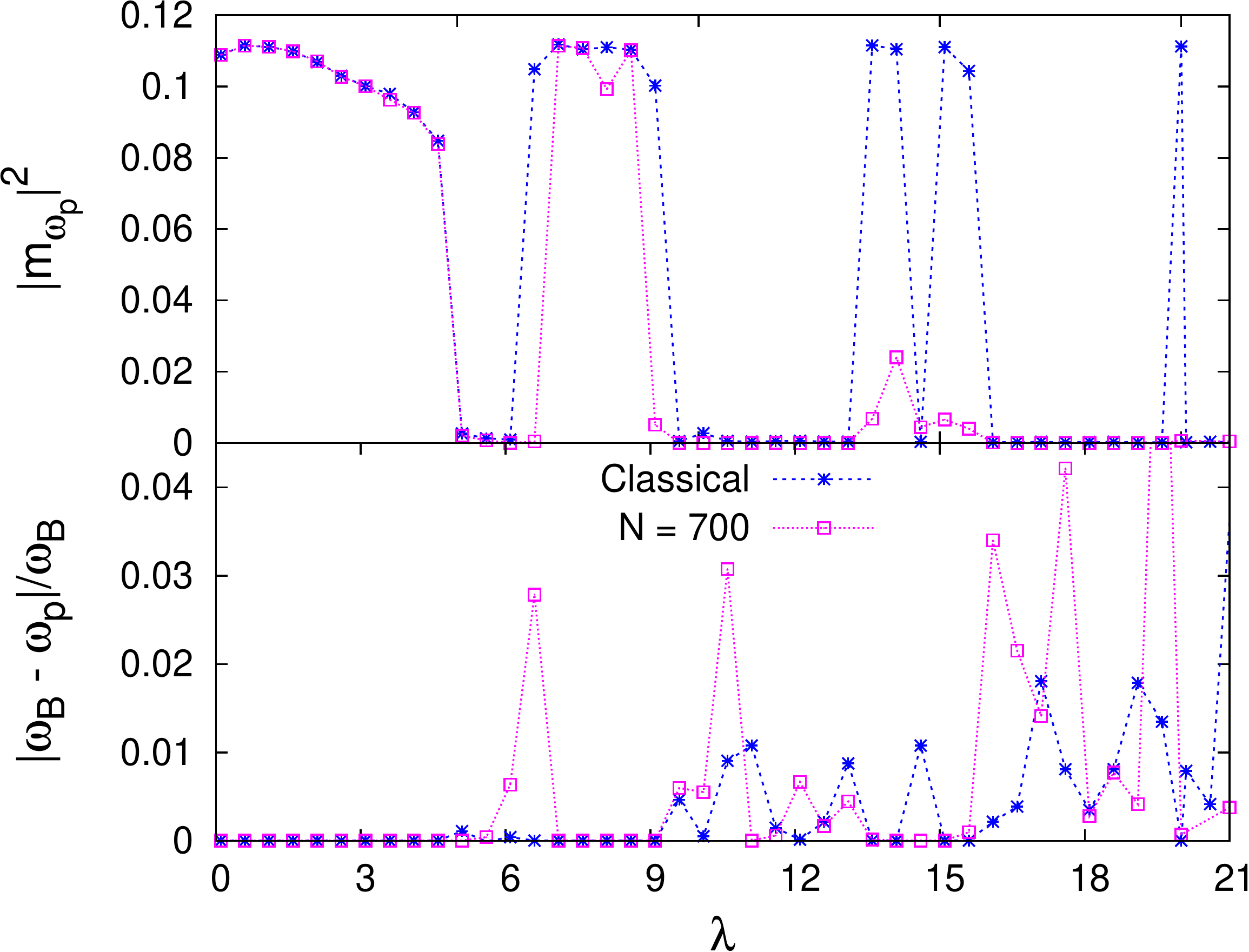}}\\
  \end{tabular}
  \caption{(Upper left panel) Dependence on $\phi$ of the height of the main peak $|m_{\omega_p}|^2$ in the power spectrum of the Fourier transform of the classical $z$-magnetization and the quantum one with $N=400$ ($\lambda=0$). (Lower left panel) Corresponding value of the frequency $\omega_p$ vs $\phi$: we can see an interval where it is locked to $\omega_B=\pi/\tau$. (Upper right panel) $|m_{\omega_p}|^2$ versus $\lambda$ in the classical and in the quantum case with $N=700$ ($\phi=\pi$). (Lower right panel) The same plot for the corresponding $\omega_p$. (Numerical parameters: $h_i = h = 0.32,\,\tau = 0.6$; the Fourier transform has been performed over $K=65536$ driving periods.)}
%
  \label{figklass}
\end{figure*}

From a quantitative point of view, we can also see a clear transition from time-translation symmetry breaking to its absence in the classical evolution of $m_n^z$.
 We can see this in the lower left panel of Fig.~\ref{ponka_phi:fig}, where we show the time-dynamics, and in the lower right panel where we plot the power spectrum of the Fourier transform: a marked peak at $\omega_B$ appears~\cite{Note2} only when there is time-translation-symmetry breaking (for $\phi=\pi$ and $\phi=0.94\pi$). We can have confirm of this looking at Fig.~\ref{figklass}: here we plot the classical height $|m_{\omega_p}^z|^2$ of the maximum peak in the $z$-magnetization transform and its frequency $\omega_p$ vs $\phi$; we also plot the same quantities in the $N=400$ case. We see an interval around $\phi=\pi$ where $\omega_p=\omega_B$ and the height depends continuously on $\phi$. This interval is consistent with the one where we see time-translation symmetry breaking for $N$ finite. We can see that inside this interval the classical value and the $N=400$ one are indistinguishable. At the boundaries of the interval, $\omega_p$ jumps away from $\omega_B$ in a way which seems discontinuous in both the finite $N$ case and the classical one. 

\begin{figure*}
  \begin{center} 
    \begin{tabular}{cc}
      \hspace{-0.5cm}\resizebox{80mm}{!}{\includegraphics{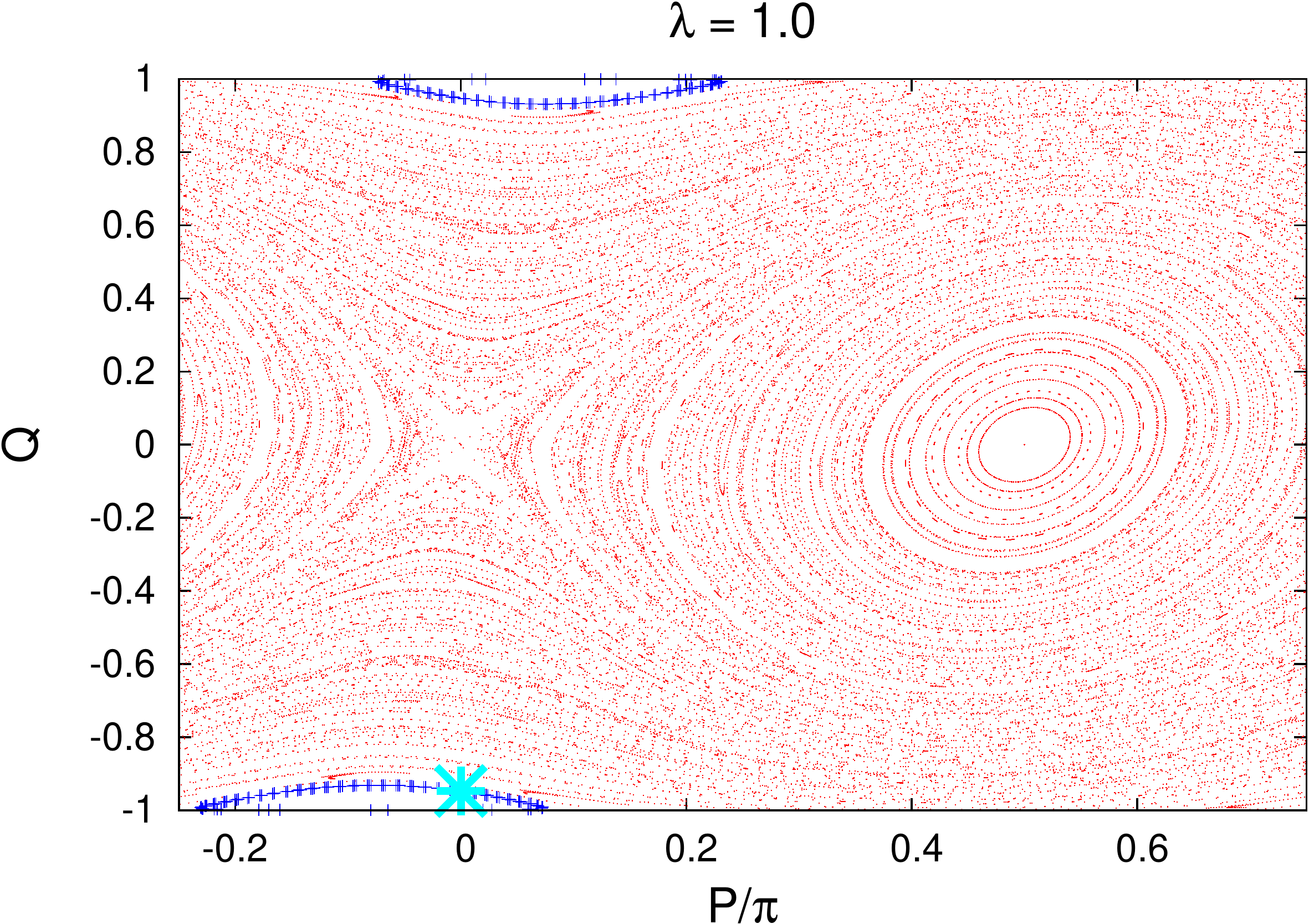}}&
      \hspace{-0.cm}\resizebox{80mm}{!}{\includegraphics{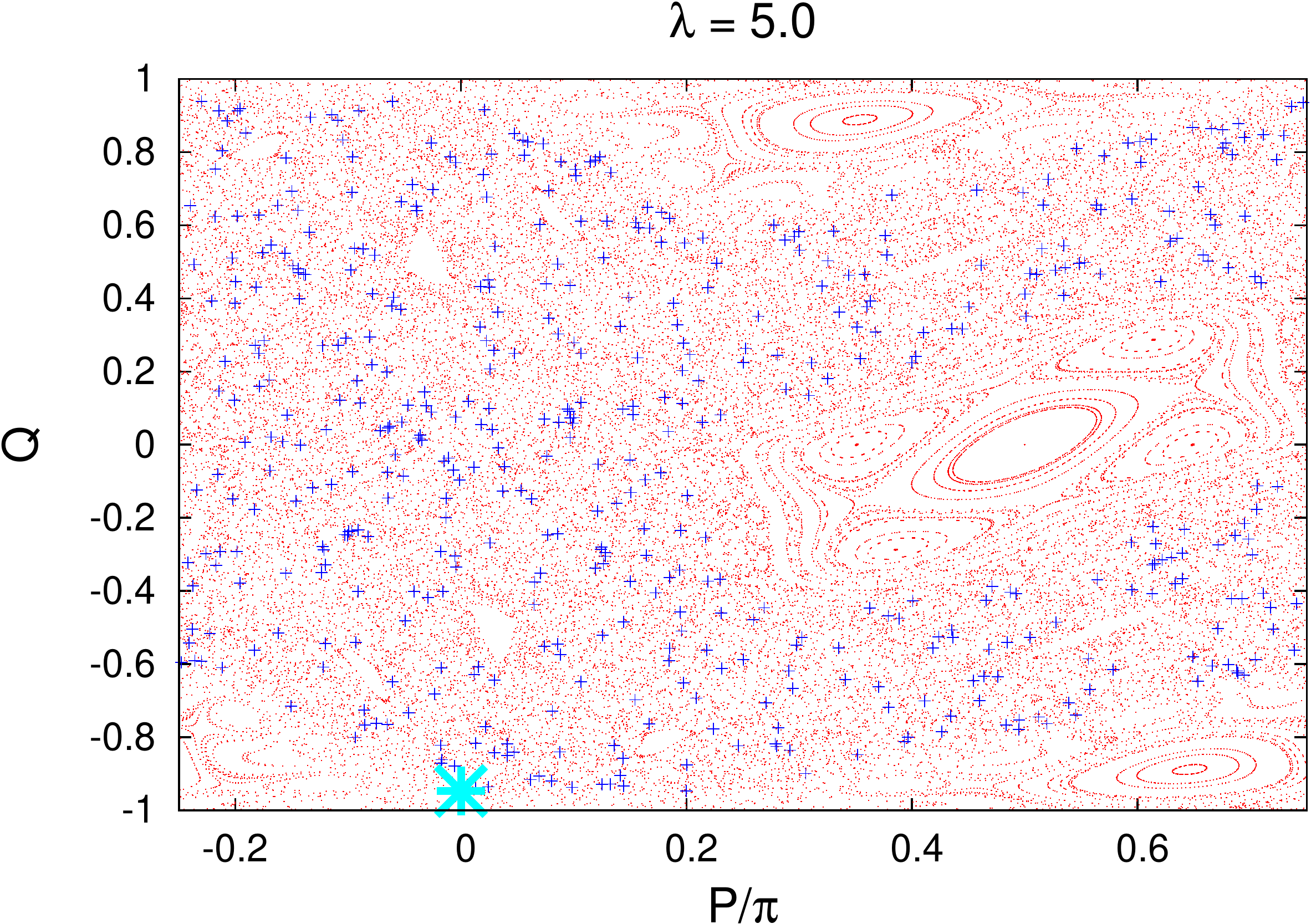}}\\
      \\
      \hspace{-0.5cm}\resizebox{80mm}{!}{\includegraphics{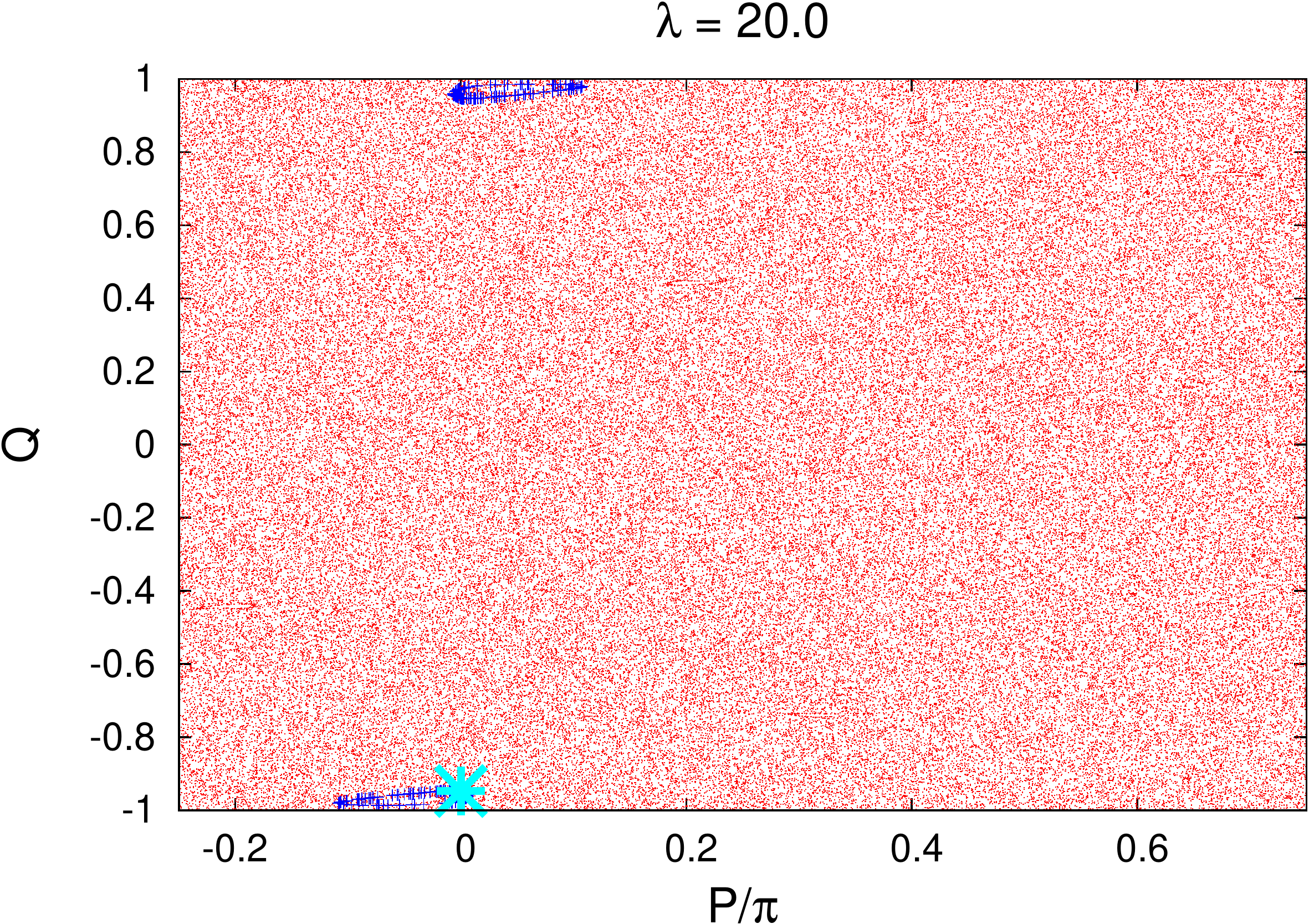}}&
      \hspace{-0.cm}\resizebox{80mm}{!}{\includegraphics{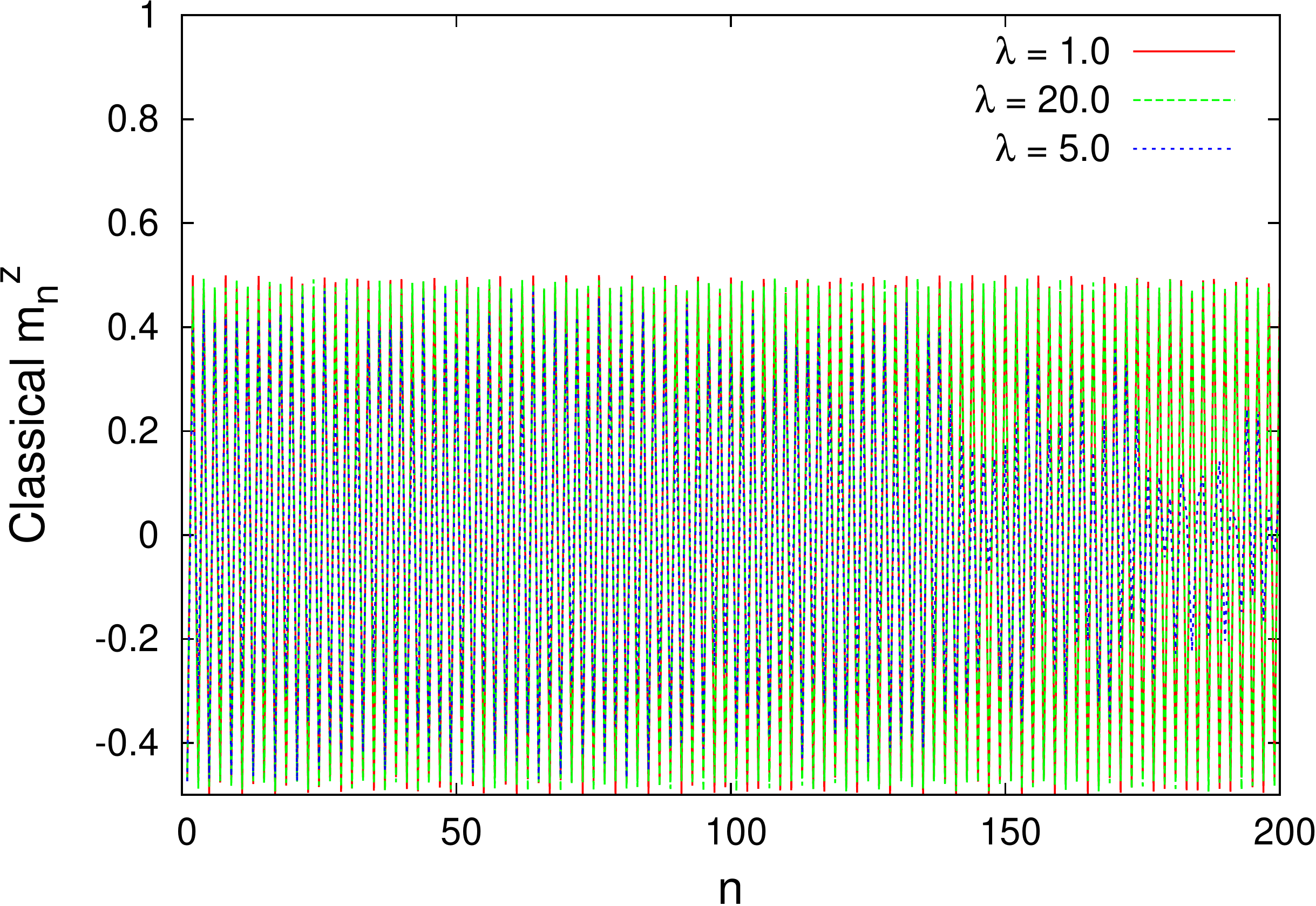}}\\
    \end{tabular}
  \end{center}
\caption{ (Upper panels) Classical phase space for different values of $\lambda$. (Lower left panel) Stroboscopic evolution in time of the classical $z$-magnetization.
(Lower right panel) The power spectrum of the corresponding Fourier transform. (Numerical parameters $h_{\rm i} = h = 0.32,\,\phi = \pi,\,\tau=0.6$. The Fourier transforms have been performed over $K=65536$ driving periods.)}
\label{ponka_lambda:fig}
\end{figure*}

In Fig.~\ref{ponka_lambda:fig} we present the results when $\phi=\pi$ and we consider different values of $\lambda$. We still plot in blue the points which represent the evolution of the initial state, chosen with $h_i=h$. The dynamics is 
increasingly chaotic when we consider larger $\lambda$. For instance, when $\lambda=1.0$ (upper left panel) the initial state is on a regular 
trajectory which has a symmetric partner in the lower part of the graph: the kicking swaps periodically one trajectory to the other and there is 
the time crystal. For larger $\lambda$, the initial state can fall inside a chaotic region and there is no period doubling ($\lambda=5.0$, upper right panel). In this case the destruction of the 
time-translation symmetry breaking is related to chaos~\cite{Russomanno_EPL15,Haake_ZPB86}. 
In the lower left panel we can see a quite interesting case: here $\lambda=20.0$ and almost all the phase space is chaotic but two small regions: one is around the initial state and the other is the symmetric one under $P\to-P$, $Q\to-Q$. The existence of these small regular regions between which the dynamics swaps at each kick (see the blue line representing the evolution of the initial state) is enough to give rise to time-translation symmetry breaking. If we had chosen initial conditions inside the chaotic region, instead, we would have seen no time-crystal behaviour. 
In the lower right panel we can see the persistence of the period-doubling oscillations for $\lambda=1$ and $\lambda=20$; they instead decay when $\lambda=5.0$.
This reflects in the power spectrum of the Fourier transform (not shown): there is a marked peak at $\omega=\omega_B$ for $\lambda=1.0$ and $\lambda=20.0$ but not for $\lambda=5.0$. 

In order to have a clearer picture, we are going to study how the properties of this peak do depend on $\lambda$. We plot the peak $|m_{\omega_p}^z|^2$ and its frequency $\omega_p$ versus $\lambda$, both for the classical and the quantum case with $N=700$, in the right panel of Fig.~\ref{figklass}. We see that the system commutes some times from the time-translation-symmetry breaking to its absence, both in the height of the peak and in its frequency being or not locked to $\omega_B$. We find that the time-crystal regions in the classical case are slightly larger than those in the $N=700$ case. 
The reason is that the quantum finite-$N$ initial state is not a point, but a wave packet whose Wigner function has width $\sim1/\sqrt{N}$ in the $Q$ and $P$ directions~\cite{Sciolla_JSTAT11,Bapst_JSTAT12}. Even if the center is on a regular trajectory (giving rise to a time crystal in the classical limit), part of the wave-packet is on chaotic trajectories which deviate exponentially fast from each other~\cite{Ott:book}. Eventually, the Wigner function gets uniformly spread all over the phase space and there is no time-translation symmetry breaking. That's why there are values of $\lambda$ (for instance $\lambda=20.0$ -- see the right panel of Fig.~\ref{figklass}) where the classical system shows a time-translation symmetry breaking, but the finite $N$ one does not. In the case $\lambda=18.5$ considered in Fig.~\ref{lambda:fig}, all the classical phase space is chaotic and the time-translation symmetry breaking is absent both for the quantum case with finite $N$ and the classical one with $N$ infinite. 


\section{Experimental realization with trapped ions} \label{experimentum:sec}
\red{The dynamics we are interested in is immediately available in linear trap ion experiments~\cite{blatt2012quantum}. In these settings, spin degrees of freedom are represented by internal states of individual ions, confined in one dimension in a Paul trap with strong transverse confinement. The interactions between ions are mediated via phonon modes, which are coupled to the internal degrees of freedom via lasers. The resulting interaction strength depends on the specifics of such coupling: in case of off-resonant coupling, the effective ion-ion interaction decays as a power of the inter-ion distance, with corresponding nearest-neighbour strength of the order of 100 Hz. An even more favourable situation is the one of infinite ranged couplings: in this case, the detuning to the motional degrees of freedom can be relatively small, thus guaranteeing stronger couplings of order 1kHz for systems of approximately 20 ions~\cite{blatt2012quantum,Jurcevic:2016aa,Neyenhuis:2016aa}. In particular, ions hosting $s=1/2$ degrees of freedom in optical qubits, such as Ca$^{+}$, are well suited to realizing spin models with infinite range interactions. This is possible not only in analogue but also using a digital simulation approach, where different interaction forms -- including the ones corresponding to the LMG model -- have already been realized (see, e.g., Ref.~\onlinecite{lanyon2011}). We refer the reader to recent works~\cite{Jurcevic:2016aa} and the review in Ref.~\onlinecite{schaetz2012,jurkevic2014} for further details.}

\red{The kicking protocol can be easily implemented by controlling the laser fields coupling internal and motional modes. In such a way, it is possible to engineer time periods where only the transverse field is present, alternated to periods displaying the full Hamiltonian dynamics, as was recently demonstrated in Ref.~\onlinecite{Neyenhuis:2016aa}. { In terms of time-scales, the main limitations are due {\it i)} to decoherence, and {\it ii)} to the fact that the switching of the different Hamiltonian parts takes place on finite time intervals, and thus on long time-scales the effective dynamics might differ with respect to the one we discuss here. The estimates for both error sources can be directly inferred from Ref.~\onlinecite{Neyenhuis:2016aa}, where results were consistent with time-crystal behaviour up to several tens of periods.} These estimated time-scales shall warrant a clear observability of the predicted time-crystal behaviour derived above.}
\section{Conclusion} \label{concolina:sec}
We have found a time-crystal behaviour in a kicked \red{infinite-range-interacting} spin chain. The fact of being \red{infinite-ranged} is crucial to have a full class of $\mathbb{Z}_2$ symmetry breaking initial states giving rise to the time-translation symmetry breaking in the ensuing evolution. We have checked the robustness of our time crystal under changing of the kicking parameters and given an interpretation in terms of the phase space properties of the classical limit of this model, attained when the number of spins goes to infinity. This analysis led us to map out the dynamical phase diagram of this model. We have then explored the properties of Floquet states and quasi-energies, showing that they have properties similar to those found for the time-crystal considered in Ref.~\onlinecite{Nayak_PRL16}. We 
remark that our findings are immediately relevant to experiments in trapped ion systems.

One possibility of future work would be to further explore what happens when interactions are long but not infinite range. In Ref.~\onlinecite{wowei_17:preprint} the authors consider disorder and power law interactions decaying with exponent 3: they are enough only for a quasi-time crystal behaviour which decays extremely slowly. In our infinite range case the decay exponent is 0: we aim to see what happens in absence of disorder in order to understand if there is a transition or a crossover to a trivial system when increasing the value of the exponent. Other perspectives of future work would be to look for the existence of time-crystal behaviour in other \red{infinite-range-interacting} models showing a standard symmetry breaking, for instance the \red{infinite-ranged} Bose-Hubbard model~\cite{Sciolla_JSTAT11} or the Dicke model~\cite{Ray_arXiv16}.

\acknowledgements{ We acknowledge useful discussions with E.~G. Dalla Torre, M.~Schir\`o, A.~Silva and especially F.~M.~Surace. We acknowledge V.~Khemani and S.~Sondhi for useful comments on the manuscript. This work was supported in part by EU-QUIC, CRF, Singapore Ministry of Education, CPR-QSYNC (R.~F.), SNS-Fondi interni 2014 (R.~F., A.~R.).}

%

%
\end{document}